\documentclass[epj]{svjour}
\def\lsim{\mathrel{\rlap{
\lower4pt\hbox{\hskip-3pt$\sim$}}
    \raise1pt\hbox{$<$}}}     %less than approx. symbol
\def\gsim{\mathrel{\rlap{
\lower4pt\hbox{\hskip-3pt$\sim$}}
    \raise1pt\hbox{$>$}}}     %greater than or approx. symbol

\usepackage{graphicx}
\begin{document}
\title{Lattice QCD Constraints on the Nuclear Equation of State}
\author{A. S.~Khvorostukhin\inst{1}%\footnote{e-mail: hvorost@theor.jinr.ru},
\and
V. V.~Skokov \inst{1,2}%\footnote{e-mail: V.Skokov@gsi.de},
\and
V. D.~Toneev \inst{1,2}%\footnote{e-mail: V.Toneev@gsi.de} 
\and
K.~Redlich \inst{2,3} %$\footnote{e-mail: redlich@ift.uni.wroc.pl}  }
}

\institute{Joint Institute for Nuclear Research,
 141980 Dubna, Moscow Region, Russia \and Gesellschaft
 f\"ur Schwerionenforschung mbH, Planckstr.$\!$ 1,
64291 Darmstadt, Germany \and University of Wroclaw, PL-50204 Wroclaw, Poland}
\date{Received: date / Revised version: date}

\abstract{
Based on the quasi-particle description of the QCD medium  at
finite temperature and density we formulate  the phenomenological
model for the equation of state that exhibits crossover or the
first order deconfinement phase transition.  The models are
constructed in such a way   to be thermodynamically consistent and
to satisfy the properties of the ground state nuclear matter
comply with constraints from intermediate heavy--ion collision
data. Our equations of states  show quite reasonable agreement
with the recent lattice findings  on temperature and baryon
chemical potential dependence of relevant thermodynamical
quantities in the parameter range covering both the hadronic and
quark--gluon sectors.  The model predictions on the isentropic
trajectories in the phase diagram are shown to be consistent with
the recent lattice results. Our  nuclear equations of states  are
to be considered  as an input to the dynamical models describing
the production  and the time evolution of a thermalized medium
created in heavy ion collisions in a  broad energy range from SIS
up to LHC.
}
\maketitle

%\today

\section{Introduction}
QCD at the finite temperature $T$ and/or baryon chemical potential
$\mu_B$ is of fundamental importance, since it describes the
relevant features of particle physics in the early universe, in
neutron stars and in heavy--ion collisions (see e.g.
\cite{W00,K02}). With the relativistic heavy ion collision
experiments at   AGS, SPS and RHIC accelerators one explores the
phase diagram of strongly interacting  matter in a broad parameter
range of temperature and baryon density. Lattice QCD results on
the Equation of State (EoS) of QCD matter provide a basic input
for the analysis of experimental signatures of a possible
quark--gluon plasma formation in heavy--ion collisions. Directly
addressing the EoS, hydrodynamics realizes the connection between
the matter properties and observables. The hydrodynamic treatment
of the whole time-evolution of colliding nuclei requires knowledge
of the\ \  nuclear\ \  EoS within a large interval of its thermodynamic
variables covering both quark--gluon and hadronic sectors.

In the recent years a  significant progress has been made in
understanding the phase diagram of QCD at non-zero baryon chemical
potential as the
 nonperturbative lattice QCD methods  were  extended to access the
relevant regions of the phase diagram. Recently, the first lattice
calculations have been performed for a non-vanishing $T$ and
$\mu_B$ for
 systems with $N_f=2$~\cite{Allton}  and $N_f=2+1$~\cite{Fodor02,Fodor04}
 flavors. However, due to a set of approximation the  Lattice Gauge
 Theory (LGT)
is still not able to provide results on the  properties of the
hadronic matter in the confined phase. LGT is also restricted to
moderate values of the baryon chemical potential $\mu_B$ such that
$\mu_B \lsim T$. That is why  different phenomenological models
are required to describe thermodynamic properties and equation of
state of QCD matter for larger baryon densities. Obviously, such
models depend on the set of parameters that are usually fixed to
reproduce existing LGT results as well as the basic
phenomenological properties of the nuclear matter obtained from
the experimental data.
 Recently, the thermodynamics of
the quark--gluon phase was  interpreted quite successfully within
the QCD inspired massive quasi-particle
models~\cite{LH98,Pesh96,Szabo03,Rebhan03,Cleymans,Bla,Weise01,%
Weise05,IST05}.
On the other hand, demonstrated by lattice calculations, a rapid
growth of the energy density $\varepsilon$ and pressure $p$ when
approaching the  critical temperature $T_c$ was  shown to be
reproduced in terms of the  hadron resonance gas model with scaled
masses \cite{KRT-1,KRT-2}. Only recently there were  attempts to
describe lattice QCD thermodynamics both above and below $T_c$ in
terms of a  field theoretical model, including features of both
deconfinement and chiral symmetry restoration~\cite{RTW05}, as
well as  within some  phenomenological models that are based on
lattice QCD  results for the quark--gluon partition
function~\cite{ADK05}. Some unique parametrization of the QCD EoS
below and above $T_c$ was also presented in Ref. \cite{BKS04}.

The phenomenological equation of state  should be not only
thermodynamically consistent~\cite{TNFNR03} but should  also be
capable to reproduce the global behavior of the nuclear matter
near the ground state and its saturation properties. In addition,
there are experimental restrictions coming from the flow analysis
in heavy--ion collisions which limit the acceptable theoretical
values of pressure in a finite interval of baryon densities $n_B$
at $T=0$~\cite{Dan02}. Some constraints on the EoS are  also
imposed through   the analysis of cold charge--neutral  baryonic
matter in $\beta$-equilibrium compact stars~\cite{IKKSTV,KB06}.
There are also  essential constraints on the model properties
coming from the recent LGT results.

In this paper, we  will  construct the EoS of a strongly
interacting QCD matter with  deconfinement phase transition that
satisfies the above mentioned  hadronic constraints and those
imposed by the recent lattice QCD results obtained  for  $(2+1)$
-- flavor system at the  finite $T$ and non-vanishing  baryonic
chemical potential.

The paper is organized as follows: In Section 2 we introduce the
quasi-particle model for  the  EoS with  deconfinement phase
transition. In Section 3 the model predictions are compared  with
the recent lattice data obtained  in  (2+1)--flavor QCD  at the
finite $T$ and $\mu_B$. Our results and comments  on the
properties of the QCD equation of state  and thermodynamics are
summarized in the last Section.

\section{The Equation of State}

Lattice  results  show that even at  temperature $T$ much larger
than deconfinement temperature   $T_c$  the thermodynamical
observables like pressure or entropy, baryon number and  energy
density are still by $\gsim 20\%$ deviating from their asymptotic
ideal gas values. Such deviations observed at $T>2T_c$ were  shown
to be well understood by a systematic contribution in the
self--consistent implementation of quasi-particle masses in the
HTL--resummed perturbative    QCD \cite{jp}. On the other hand,
the LGT thermodynamics below $T_c$ was  shown to be well
reproduced by the hadron resonance gas  partition function
\cite{KRT-1,KRT-2}. To possibly describe the thermodynamics at
$T=T_c$ or near the phase transition an additional model
assumptions are required \cite{quasi,Szabo03}.

It is clear from the above that the  straightforward model for the
QCD EoS can be constructed by   connecting a  non-interacting
hadron resonance gas in the low temperature phase with an ideal
quark gluon--plasma in some  non-perturbative bag   for the color
deconfined phase~\cite{Cleymans}. These phases are matched at the
phase transition boundary by means of the Gibbs phase equilibrium
condition. By construction, this approach yields the  first order
phase transition. Such MIT bag-like model~\cite{CJJT74} is so far
the simplest  method  to implement the confinement phenomenon in
the EoS, though it has some serious shortcomings.

A more complete method to model QCD EoS is based on the effective
Hamiltonian that includes interactions of the constituents. In
  the quasi-particle approximation such
Hamiltonian can be modelled   through density--dependent
mean--field interactions ~\cite{TNFNR03,NST98,TNS98}:
\begin{eqnarray}
%\begin{multline}
 H &&= \sum_{j\in h,q,g} \sum_s \int d{\bf r} \ \psi^+_j({\bf r},s) \ \nonumber
\cdot \\ &&\left( \ \sqrt{-\nabla^2 + m^2_i}+U_j(\rho) \ \right) \
\psi_j({\bf r},s) - C(\rho ) \cdot V \ , \label{eqH} 
%\end{multline}
\end{eqnarray}
where $j$ \ \ enumerates \ \ different species of quasi-particles (hadrons
and/or unbound quarks and  gluons) and  $s$ stands for their
internal degrees of freedom. Here $U_j(\rho )$ is the
density-dependent mean--field acting on the quasi-particle $j$
described by the field operator $\psi_j $ with $m_j$ being the
current mass of quarks and gluons or the free mass of  hadrons.
Applying  the  density-dependent Hamiltonian (\ref{eqH}) in the
partition function requires some additional  constraints that are
needed  to fulfill the thermodynamic consistency
condition~\cite{shan}~:
\begin{equation}
\langle \frac{\partial H}{\partial T} \rangle\, = \,0\, , \quad
\langle \frac{\partial H}{\partial \rho_{j}} \rangle\, = \,0 \;\;
, \label{eq2}
\end{equation}
where $\langle A \rangle$ denotes the average value of the
operator $A$ over the statistical ensemble. With the Hamiltonian
(\ref{eqH}) the conditions (\ref{eq2}) can be  also expressed as
~\cite{shan}
\begin{eqnarray}
\sum\limits_{j}\;\rho_{j}\,
         \frac{\partial U_{j}}{\partial \rho_{i}}\; - \;
                         \frac{\partial C}{\partial \rho_{i}}
                                                          \;=\;0
                                                \,,\, \,
\sum\limits_{j}\;\rho_{j}\,
             \frac{\partial U_{j}}{\partial T}\;-\;
                     \frac{\partial C}{\partial T}\;=\;0\,.
\label{eq3}\end{eqnarray}

It can be shown ~\cite{TNFNR03,NST98,TNS98} that  the conditions
(\ref{eq3}) are satisfied only if the mean field $U_j(\rho)$ and
the correcting function $C(\rho )$ are
 temperature independent.

In  the following, we consider the basic structure of the
effective Hamiltonian (\ref{eqH}) to model the EoS of hadronic and
quark--gluon plasma phase.

\subsection{The hadronic phase}

The hadronic phase is considered as a gas of hadrons and
resonances in the thermodynamic equilibrium. In general, the
particle density of species $j$ is obtained from
\begin{eqnarray}
n_j&\equiv& n_j(T,\mu_j-U_j )=\nonumber \\ &&\frac{d_j}{2\pi^2}\int_0^{\infty} dk\
k^2 \ f_j(k,T,\mu_j-U_j)~, \label{eqt1}
\end{eqnarray}
where the one-particle distribution function with an argument $z$
is
\begin{eqnarray}
 f_j(k,T,z) = \left[ \  \exp \left( \frac{\sqrt{k^2 +m_j^2}-z}{T} \right)
 + {\cal L}_j \right]^{-1}
\label{eqt2}
\end{eqnarray}
with ${\cal L}_j=+1$  for fermions and  ${\cal L}_j=-1$ for
bosons, while $d_j$ is the spin--isospin degeneracy factor. The
chemical potential $\mu_j$ is related to the baryon ($\mu_B$) and
strangeness ($\mu_S$) chemical potentials
\begin{equation}
\mu_j = b_j \ \mu_B+s_j \ \mu_S\,, \label{eqt1a}
\end{equation}
with $b_j$ and $s_j$ being  the baryon number and strangeness of
the particle $j$. The hadronic potential $U_{j} \equiv
U_{j}^{(h)}$ is described by a non-linear mean--field
mo\-del~\cite{Zim}
\begin{eqnarray}
U_{j}^{(h)}\;=  g_{r,j}\;\varphi_1 (x) + g_{a,j}\;\varphi_2 (y)\;,
\label{eqZ}
\end{eqnarray}
where $g_{r,j} > 0$ and $g_{a,j} < 0$ are repulsive and attractive
coupling constants, respectively. The effect of interactions
results also in an  additional density-dependent term $C(\rho)$
that contributes  to the thermodynamic  pressure and energy
densities. If the particle interaction is  taken in  form of
(\ref{eqH}), the thermodynamic consistency implies that the
functions $\varphi_1(x)$ and $\varphi_2(y)$ depend only on
particle densities. In Ref. \cite{Zim} these functions were chosen
such that
\begin{eqnarray} b_1
\varphi_1 = x, \quad -b_1 (\varphi_2 + b_2 \varphi_2^3 ) = y
\label{eq22}
\end{eqnarray}
where
$$ x=\sum\limits_{i} g_{r,i}\; n_{i},\quad
y=\sum\limits_{i} g_{a,i}\;n_{i}\;. $$ with   $b_1$ and $b_2$
being  free parameters. The $\varphi_2^3$ term is introduced to
get a slower than linear increase of attraction with density at a
high compression   as it happens in the relativistic mean--field
models. Having in mind that the the hadronic EoS will be compared
with that of the quark--gluon plasma, it is convenient to rewrite
(\ref{eq22}) in terms of a number of constituent quarks and
antiquarks  $\nu_i$~:
\begin{eqnarray}
\rho_j=\nu_j n_j \equiv \nu_j n_j(T,\mu_j-U_j)~.
 \label{rhoj}
\end{eqnarray}

In the original paper \cite{Zim}, the hadronic phase was modelled
as a mixture of  nucleons and $\Delta$'s (i.e. $j=N,\Delta$).
Following~\cite{TNFNR03}, we generalize this approach by including
all hadrons and resonances with the mass  up to  1.6 GeV. One also
assumes that all coupling constants scale with the number of
constituent quarks~\cite{TNFNR03}~:
\begin{eqnarray}
U_{j}^{(h)} = \nu_j\,\Bigl(
[\widetilde\varphi_1(\rho^{(h)})]^\alpha +
\widetilde\varphi_2(\rho^{(h)})\,\Bigr)\;, \label{eq23}
\end{eqnarray}
where $\widetilde\varphi_1$ and $\widetilde\varphi_2$ satisfy Eq.
(\ref{eq22}) in the following form
\begin{eqnarray}
 c_1 \widetilde\varphi_1^{\alpha} = \rho^{(h)}, \quad
-c_2 \widetilde\varphi_2 - c_3 \widetilde\varphi_2^3 =
\rho^{(h)}\; \label{eq24}
\end{eqnarray}
with $\rho^{(h)}=\sum\nolimits_{j}\nu_j \rho_{j}=3\sum_B
n_j+2\sum_M n_j$. As  compared to Eq. (\ref{eq22}) a free
parameter $\alpha$ is also introduced in Eq. (\ref{eq23}). This
parameter is used to control the strength of the repulsive
interactions at high density ~\cite{NST98,TNS98}. The parameters
in Eq. (\ref{eq24}) are expressed as  \cite{NST98}:
$$ c_1
= \frac{b_1}{(g_{r,j}/\nu_j)^2}, \quad c_2 =
\frac{b_1}{(g_{a,j}/\nu_j)^2}, \quad c_3 = \frac{b_1
b_2}{(g_{a,j}/\nu_j)^4}\; $$ and are fixed by  requiring that the
properties of the ground state ($T=0$ and
$n_B=n_0\approx0.15\;{\rm fm^{-3}})$ of the nuclear matter are
reproduced: zero pressure,  binding energy per nucleon of -16 MeV
and incompressibility of 210 MeV.

Solving the cubic equation (\ref{eq24}), one gets  the
interaction potential as
\begin{eqnarray}
U_{j}^{(h)}\;= \nu_j \left[ \frac{1}{c_1} \cdot
(\rho^{(h)})^\alpha - F(\rho^{(h)} ) \right]
 \label{eqZ1}
\end{eqnarray}
where the function $F$
 depends  on the  density of {  quarks bounded inside hadrons} as
 follows
% (to be a real solution of the cubic equation
%(\ref{eq24}) as given in~\cite{KK68} )
\begin{eqnarray}
\label{Ff} F(t) = \frac{12^{1/3}}{6} \eta - 2 \beta \eta^{-1} \ 
\mbox{with} \ \eta = \left( \frac{t}{a} + \sqrt{\beta^3 +
\frac{t^2}{a^2}} \right)^{\frac{1}{3}}.
\end{eqnarray}
Here $a, \beta$ are proportional to the coefficients of the
equation (\ref{eq24}): $a =c_3/9$  and $\beta
=c_2/12^{1/3}$.

 In this  representation we obtained  for the hadronic
pressure
 \begin{eqnarray}
&&p^{(H)} (T,\mu_j-U_j^{(h)} )=\label{EoS:eqp} \\ \nonumber \sum_{j\in h} &&
 \frac{d_j}{6\pi^2}   \int_0^{\infty}
 \frac{k^2}{ \sqrt{k^2 + m_j^2} } f_j(k, T, \mu_j-U_j^{(h)})   \  k^2 dk + C(\rho^{(h)}) 
\end{eqnarray}
and for the energy density
\begin{eqnarray}
&&\varepsilon^{(H)}(T,\mu_j-U_j^{(h)} ) = \nonumber  \sum_{j\in h} \frac{ d_j}{2\pi^2}\int_0^{\infty}  \left( \sqrt{k^2 + m_j^2} + U^{(h)}_{j} \right) \cdot\\&&\cdot
f_j(k, T, \mu_j-U_j^{(h)})    k^2 dk  -C(\rho^{(h)})~,
\label{EoS:eqeps}
\end{eqnarray}
where the function $C$ is obtained from
\begin{eqnarray}
&&C(\rho^{(h)}) = \frac{1}{c_1}  \frac{\alpha}{\alpha +1 } \
(\rho^{(h)})^\alpha  - \rho^{(h)} F(\rho^{(h)}) +\\ \nonumber
&&+\int_0^{\rho^{(h)}} F(t) \ dt. \label{EoS:B_analitic}
\end{eqnarray}

\begin{figure}[thb]
\centerline{
\includegraphics[width=60mm,clip]{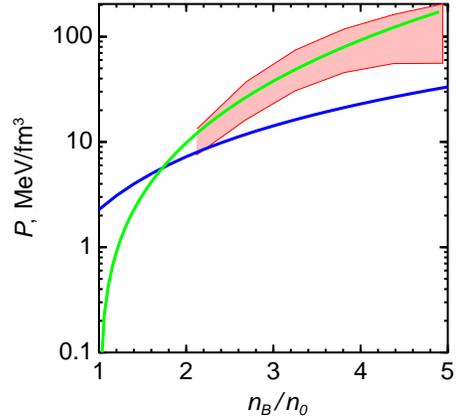}}
\caption{ Pressure as a function of baryon density for $T=0$.
 Grey and black solid lines are calculated for the modified Zimanyi
model and ideal gas EoS, respectively. The shaded region
corresponds to the Danielewicz { et. al.} constraint~\cite{Dan02}.
}
 \label{dan}
\end{figure}

\begin{figure*}[thb]
\centerline{
\includegraphics[width=60mm,clip]{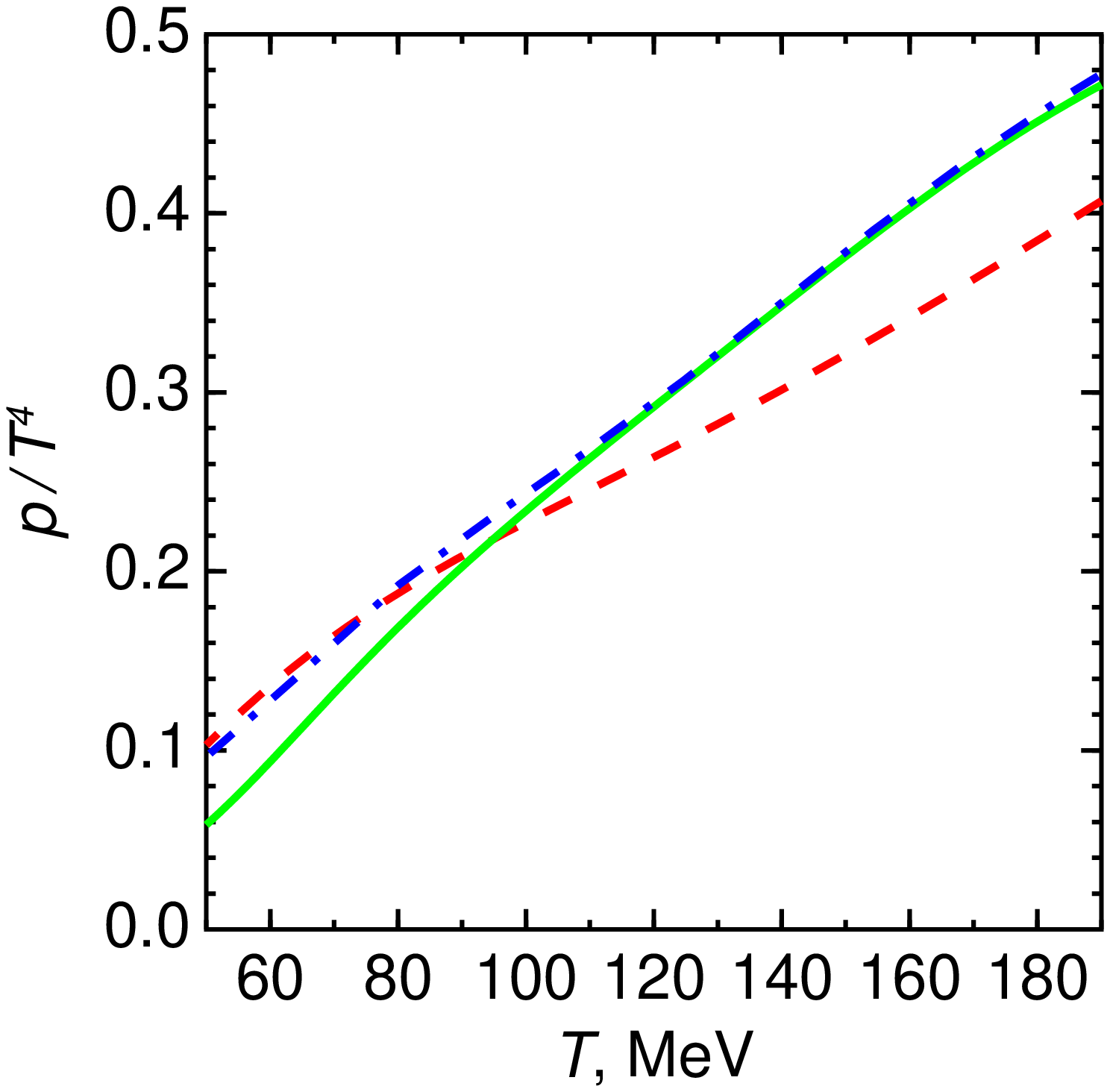} \hspace*{3mm}
\includegraphics[width=60mm,clip]{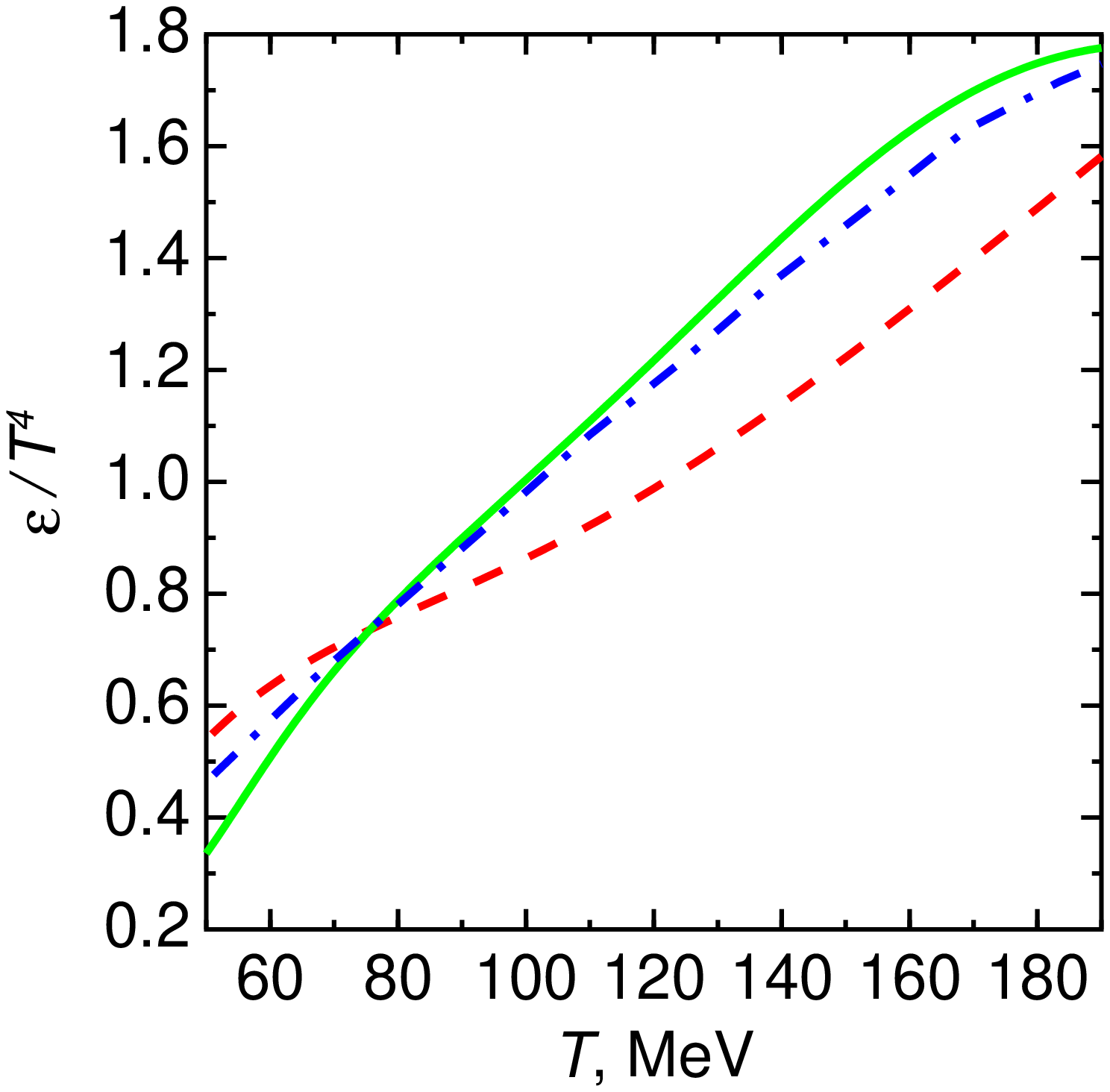}}
\caption{Temperature dependence of the reduced pressure and energy
density for an interacting pion gas ($\pi +\rho$ system). The
solid line is our result, dash-dotted  and dashed ones are the
interacting ~\cite{RW96} and ideal pion gas, respectively. }
 \label{pion_gas}
\end{figure*}

As shown in Fig. \ref{dan}, the above  hadronic EoS satisfies the
constraint resulting  from the nucleon flow analysis of heavy ion
collisions in the energy range $\lsim 10$ AGeV. The upper boundary
of the shaded area is consistent with the constraint coming  from
the analysis of the neutron star properties~\cite{KB06}. In the
high temperature regime there is also a reasonable agreement of
our model  with the thermodynamics of the interacting pion gas
from ~\cite{RW96} (see Fig. \ref{pion_gas}).

\subsection{The two--phase bag model}

In the MIT bag--like  model,  the deconfinement phase transition
is determined by matching the EoS of an ideal  relativistic gas of
hadrons and resonances to that of an ideal gas of quarks and
gluons. In the following we consider the  two--phase (2P) model
that accounts for interactions separately in the hadronic and
quark--gluon plasma phase. The hadronic phase is described within
the phenomenological mean--field model introduced  in the previous
Section. Following Eq. (\ref{eqt1}), the total baryon density and
the strangeness density in the hadronic phase can be expressed as
\begin{eqnarray}
\label{eqt1b}
n_B^H &=&  \sum_{j \in h} b_j \  n_j(T,\mu_j-U^{(h)}_j)~, \\
n_S^H &=&  \sum_{j \in h} s_j \  n_j(T,\mu_j-U^{(h)}_j)~,
\label{eqt1c}
\end{eqnarray}
where the sum is taken over all hadrons and resonances. Similarly,
the pressure and energy density of the species $j$ are given by
Eqs. (\ref{EoS:eqp}) and (\ref{EoS:eqeps}).

In the quasi-particle approximation, the QGP phase is commonly
described as a gas of partons (non--interacting point-like quarks,
antiquarks and gluons) confined in a "bag". The non--perturbative
effects associated with confinement are presented by the constant
vacuum energy $B$. The recent LGT results show that such an
approach is not adequate as the EoS differs from the asymptotic
ideal gas values even at temperatures
 as high as $100
\ T_c$~\cite{ABPS02}.  The perturbative QCD results can be,
however, improved  through  the so-called Hard Thermal Loop (HTL)
expansion. According to the HTL perturbative expansion,  the QCD
thermodynamics at high temperature is controlled by
quasi-particles\ \  with\ \  a \ \ temperature\ \  dependent mass $m_q(T)$. For
$\mu_B=0$ one gets~\cite{ABPS02,HTL}:
\begin{eqnarray}
m_q^2(T) - m_{q0}^2 = \frac{N_g}{16 Nc} T^2 g^2~. \label{htl1}
\end{eqnarray}

To model the HTL results within the mean--field approach one
introduces the  quark and gluon potentials to reproduce the
behavior of the HTL masses (\ref{htl1}) in the high temperature
limit. This in general  results in an additional equation for the
unknown gluon density. To simplify the problem   we modify the
potential so  that it coincides only with the HTL expression for
quarks. In the high temperature limit and having in mind that
$\rho \sim T^{3}$, the simplest  phenomenological choice of the
potential is
\begin{equation}
U^{(pl)} =  {\cal B} \ (\rho^{(pl)})^{1/3}, \label{mod}
\end{equation}
where ${\cal B}$ is obtained  comparing the asymptotic expansion
of (\ref{mod}) with the HTP result
\begin{eqnarray}
{\cal B} = g  \frac{\sqrt{\frac{N_g}{16 N_c}}}{ \left(
\frac{\zeta(3)}{2 \pi^2} \left( 2 d_g + 3 N_f d_q \right) \right)^
{1/3}~} \label{B}
\end{eqnarray}
with $d_q$ and $d_g$ being  the degeneracy factors for quarks and
gluons, respectively. For  $N_c = 3$ and  $N_g = 8$ one gets
\begin{eqnarray}
{\cal B}(N_f=3) = 0.2351\ g ~, \\
{\cal B}(N_f=2) = 0.2542\ g~,
 \label{Bnum}
\end{eqnarray}
 where \ \ the \ \  strong interaction coupling constant $g$ is
treated  as a free parameter.

The \ \ thermodynamic \ \ self-consistency conditions require that the
mean--field contribution to  the pressure and energy  density in
equations like Eqs. (\ref{EoS:eqp}) and (\ref{EoS:eqeps}) is
respectively
\begin{eqnarray}
 U_i^{(pl)} &=& {\cal B} \ \nu_i \ (\rho^{(pl)})^{1/3},\label{U2P}\\
C(\rho^{(pl)})&=&  \frac{\cal B}{4} \ (\rho^{(pl)})^{4/3} +B~,
\label{C2P}
\end{eqnarray}
where the plasma particle density $\rho^{(pl)}=\sum_{j \in
g,q,\bar q} \rho_j$ and the bag constant $B$ is included in  the
correcting function $C$.

With such  mean--field potentials the pressure and energy density
in the plasma phase carried  by $u,d$ and $ s$ quarks and
antiquarks is obtained as
\begin{eqnarray}
p^{Q}(T,\mu_j-U_j^{(pl)}) &=& \sum_{j \in g,q,\bar q} \nonumber
p_j(T,\mu_j-U_j^{(pl)}) -\\&&-C(\rho^{(pl)})~,
\label{eqt8} \\
 \varepsilon^{Q}(T,\mu_j-U_j^{(pl)}) &=&
\sum_{j \in g,q,\bar q} \nonumber
\varepsilon_j(T,\mu_j-U_j^{(pl)})+\\&&+C(\rho^{(pl)})~, \label{eqt10}
\end{eqnarray}
To quantify these observables we use the quark masses $m_u=m_d=65$
MeV and $m_s=135$ MeV, the gluon mass $m_g=700$ MeV  and the bag
constant $B^{1/4}={\rm 207 \ MeV}$. Such parameters \, \, \,   yield  \, a
transition temperature $T_c \approx 170 \ \mbox{MeV}$ in agreement
with the recent lattice result obtained for the vanishing net
baryon number ~ \cite{K02}.
 For  massless  gluons the   equation of state has a simple form
\begin{equation}
 p_g(T) = \frac{d_g \pi^2}{90} T^4~, \ \ \  \varepsilon_g(T) = 3  p_g(T) =
 \frac{d_g \pi^2}{30} T^4
 \label{eqt9}
\end{equation}
with $d_g$=16~.

The   baryon number and strangeness density in the quark--gluon
plasma  are obtained following  Eqs. (\ref{eqt1b}) and
(\ref{eqt1c}) from
\begin{eqnarray}
\label{eqt11b}
n_B^Q &=&  \sum_{j \in g,q,\bar q} b_j \  n_j(T,\mu_j-U_j^{(pl)} )~, \\
n_S^Q &=&  \sum_{j \in g,q,\bar q} s_j \ n_j(T,\mu_j-U_j^{(pl)}
)~. \label{eqt11c}
\end{eqnarray}

The  equilibrium between the plasma and the hadronic phase is
determined by the Gibbs conditions for the thermal ($T^Q=T^H$),
mechanical ($p^Q=p^H$) and chemical ($\mu_B^Q =\mu_B^H, \ \mu_S^Q
=\mu_S^H$) equilibrium.
 At a given temperature $T$ and baryon chemical
potential $\mu_B$ the strange chemical potential $\mu_S$ is
obtained  by requiring that the net strangeness of the total
system vanishes. Consequently, the phase equilibrium condition and
strangeness conservation imply that:
\begin{eqnarray}
\label{eqt12a}
&&p^H(T,\mu_j-U_j^{(h)}) = p^Q(T,\mu_j-U_j^{(pl)}), \\
\label{eqt12b} &&n_B =  (1-
\lambda )n_B^H(T,\mu_j-U_j^{(pl)} ) +\\ \nonumber && \quad \quad  + \lambda n_B^Q(T,\mu_j-U_j^{(pl)} ), \\
&&0=  (1-\lambda ) 
n_S^H(T,\mu_j-U_j^{(pl)}) + \\ \nonumber && \quad \quad + \lambda  n_S^Q(T,\mu_j-U_j^{(pl)} ) , \label{eqt12c}
\end{eqnarray}
where $\lambda = V_Q / V$ is the fraction of the volume occupied
by the plasma phase. The phase boundaries of the coexistence
region are found by putting $\lambda = 0$ for the hadron phase
boundary and $\lambda = 1$ for the plasma boundary. By
construction the 2P EoS results in the first-order phase
transition with discontinuous behavior of energy and baryon
densities.

According to the Gibbs  condition ~\cite{LL}, the number of
thermodynamic degrees of freedom that may be varied without
destroying the equilibrium of a mixture of $r$ phases  with $n_c$
conserved charges is ${ \cal N}=n_c+2-r$. For the considered
hadron--quark deconfinement transition $r=2$. If the baryon number
is the only conserved quantity then $n_c=1$ and ${\cal N}=1$.
Thus, the phase boundary is one--dimensional, i.e. a line. The
Maxwell construction for the first-order phase transition
corresponds   to $r=2$ and $n_c=1$. When both the baryon number
and strangeness are conserved, that is when $n_c=2$, one has
${\cal N}=2$ and therefore the phase boundary is a surface. In
such a system, a standard Maxwell construction is no longer
possible~\cite{Glend92,DCG06}\footnote[1]{In~\cite{Glend92,DCG06} the baryonic and electric charge
conservation was considered in application to a nuclear liquid-gas
phase transition. As to strangeness conservation the emphasis was
made mainly on the strangeness distillation effect~\cite{GKS93}.
Phase boundaries for this case were studied in detail
in~\cite{LH93}. More complete list of appropriate references can
be found in~\cite{TNFNR03}. }.

When two phases coexist, the system  is in general { not
homogeneous} because the phases    occupy separate domains in
space. We do not, however,  explicitly account for such a domain
structure or
 a possible surface energy contribution to the equation of state.
The only consequence of the phase separation in the considered 2P
model  is that the interaction between quasi-particles in the
plasma and hadronic phase are neglected. This is different from
the statistical mixed phase model that will be  discussed in the
next subsection.

\begin{figure*}[thb]
\centerline{
\includegraphics[width=60mm,clip]{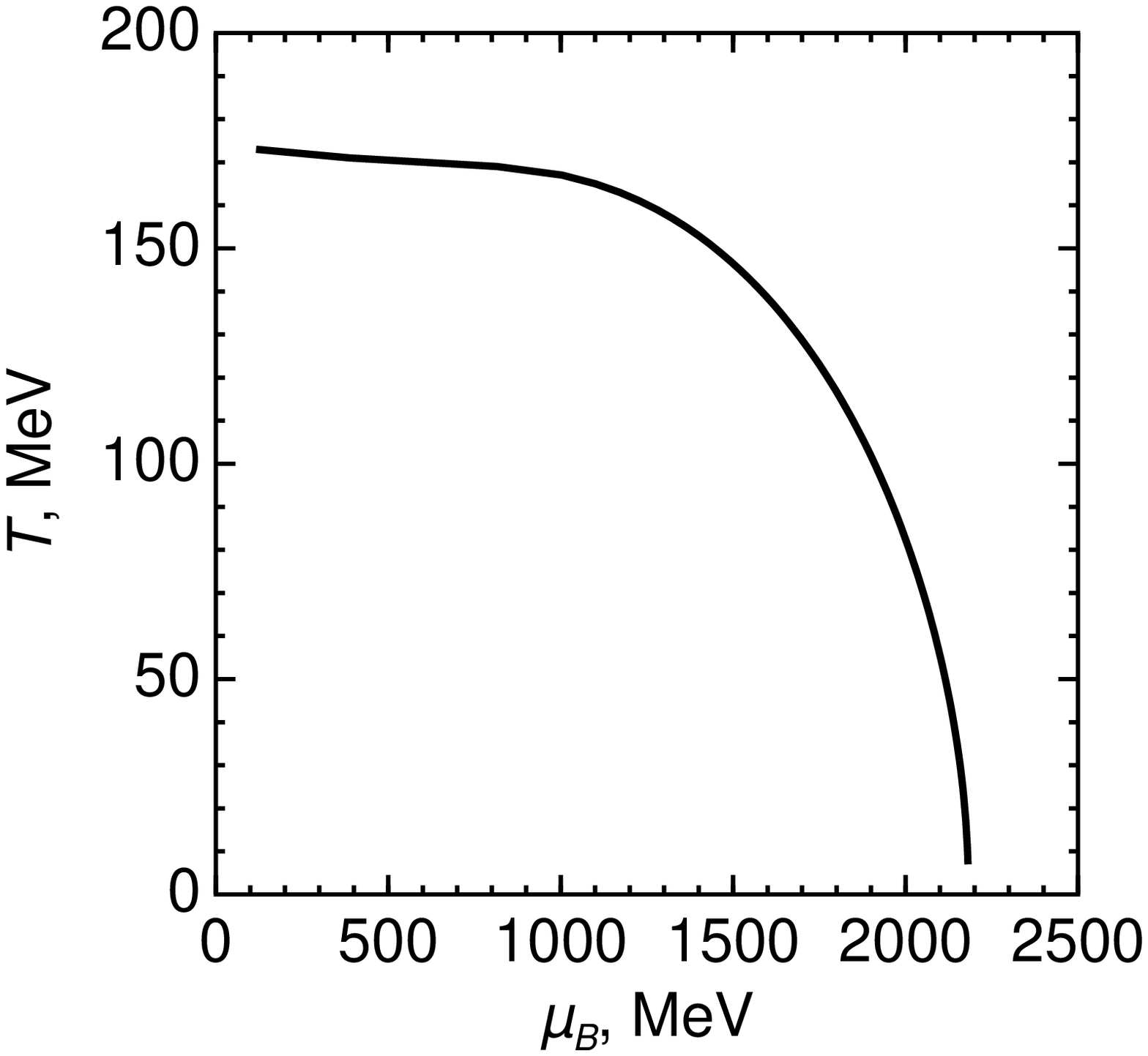} \hspace*{3mm}
\includegraphics[width=60mm,clip]{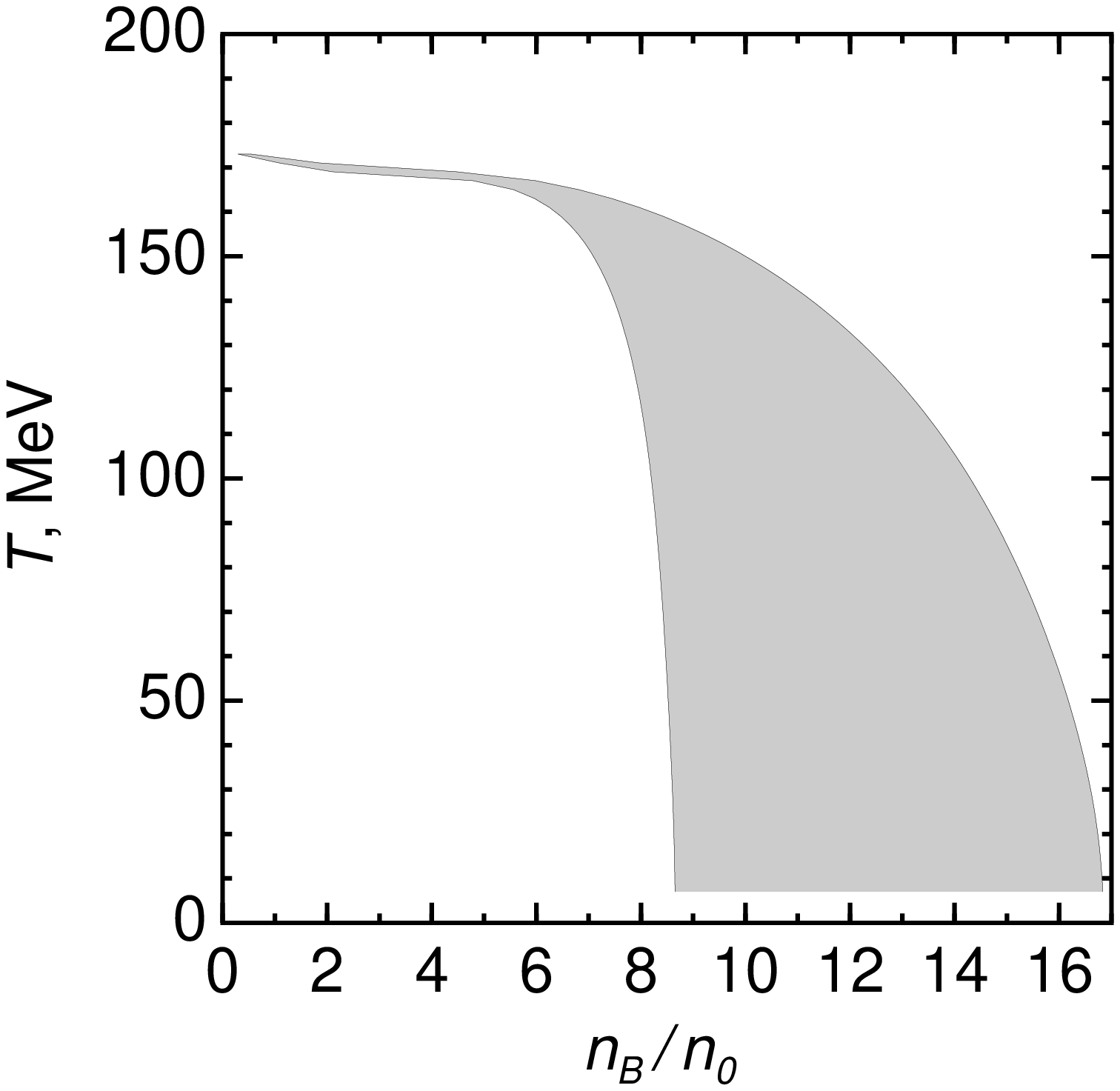}}
\caption{ The phase boundary calculated  in the 2P  model with the
physical values of parameters as explained in the next Section. }
 \label{pr}
\end{figure*}

The resulting phase boundaries between the hadronic phase and the
quark--gluon plasma  in the 2P model are shown in Fig.~\ref{pr}.
At $T=0$ the coexisting region appears at $n_B/n_0 \simeq 8$. This
 density is by factor two larger than that one obtained in the conventional
MIT-bag like  model ~\cite{TNFNR03} (see also Ref.~\cite{IKKSTV}).
This is because in our calculations the quarks and gluons are
treated as massive quasi-particles. As it will be shown in the
next Section, the finite mass of quasi-particles is needed in  the
quark gluon plasma to get the EoS that is consistent with LGT
results.

\subsection{The mixed-phase model}

In the 2P model the interactions between quark, gluons and hadrons
are entirely neglected in the coexistence region. In the following
we introduce the MP model where such interactions are possible.
The underlying  assumption of the MP model \cite{NST98,TNS98} is
that  unbound quarks and gluons { may coexist} with hadrons
forming a { homogeneous} quark/gluon--hadron phase. Since the mean
distance between hadrons and quarks or gluons in this mixed phase
may be of the same order as that between hadrons, the interaction
between all these constituents
 (unbound quarks, gluons and hadrons) plays an important
role as it  defines the order of the phase transition.

 Under a  quite general requirement for  the   confinement
of color charges, the mean--field potential of  quarks and gluons
in the  plasma phase is approximated as
\begin{eqnarray}
U_q(\rho^{(pl)})=U_g(\rho^{(pl)})&=&\\ \nonumber  &=&{{\cal
A}\over(\rho^{(pl)})^{\gamma}}+{\cal B} \ (\rho^{(pl)})^{1/3}~;
 \ \ \ \gamma >0~,
\label{eq6}
   \end{eqnarray}
where $\rho^{(pl)}=n_q+n_{\bar q} +n_g$. The second term in
Eq. ({\ref{eq6}}) is introduced to  account for  the growth of the
quasi-particle mass with the density as that obtained in  the HTL
approximation (see Eq. (\ref{U2P})). The first term in
 Eq. (\ref{eq6}) reflects  two important limits of the QCD
interactions. For $\rho^{(pl)} \to 0$, this  potential term
approaches the infinity, {\em i.e.} an infinite energy is
necessary to create an isolated quark or gluon that corresponds to
the confinement of color objects. The other extreme limit of
infinite density  is consistent with the asymptotic freedom.

The generalization of the mean--field potential from Eq.
(\ref{eq6}) to  the case of the  { mixed} quark--hadron phase is
obtained replacing $\rho^{(pl)}$ in  Eq. (\ref{eq6}) by the total
density of quarks and gluons $\rho^{(mp)}$ with
\begin{equation}
\rho^{(mp)}=\rho_q + \rho_{\bar q} +\rho_g +\eta
\sum\limits_{j}\;\nu_j\rho_{j} \equiv \rho^{(pl)}+\eta \
\rho^{(h)}~, \label{rhomp}
\end{equation}
  The presence of the total number density $\rho^{(mp)}$
in Eq. (\ref{eq6}) implies interactions between all components of
the mixed phase. For  $\eta=0$ there is no  interaction between
hadrons and unbound quarks and gluons. This case  corresponds to
such a  strong binding of hadron constituents that the presence of
free color charges in their surrounding does not result in their
color polarization, i.e. hadrons remain color neutral and do not
see quarks and gluons outside hadron. Thermodynamically,   the
potential with $\eta=0$ implies   the   first order phase
transition.  For $\eta=1$ there is  a very strong color
polarization of hadrons. Consequently, there is no difference
between bound and unbound quarks and gluons. This approximation
was used in \cite{NST98,TNS98}. Here we consider  $\eta$ as a free
parameter that is  chosen  in a way  to reproduce the LGT results
for the QCD equation of state.

 The hadronic potential in the Hamiltonian
(\ref{eqH}) was  described  by a non-linear mean--field model.
However, the presence of unbound quarks and gluons will modify
this \ \ hadronic interaction\ \  due to \ \ the polarization of   color
charges. Thus, in general
\begin{eqnarray}
U_{j}^{(mp)}=U_{j}^{(h)}+U_{j}^{(h-pl)}\;. \label{eq43}
\end{eqnarray}

The  constraints imposed by the thermodynamic consistency
conditions (\ref{eq3}) can be used   to find the  potential  for
the interaction of unbound quarks/gluons with  hadrons as
~\cite{NST98,TNS98}
\begin{eqnarray}
U_{j}^{(h-pl)}\;&=&\;\nu_j \eta \left( \frac{\cal
A}{(\rho^{(mp)})^{\gamma}} -\frac{\cal A}{(\eta
\rho^{(h)})^{\gamma}} + \right. \nonumber \\  &&+\left.{\cal B} \ [(\rho^{mp)})^{1/3}-(\eta
\rho^{(h)})^{1/3}]\right)~. \label{eq17}
\end{eqnarray}

Consequently,  the pressure and the energy density in the MP model
are obtained from
\begin{eqnarray}
&&p^{MP}(T,\mu_j-U_j^{(mp)}) = \sum_{j \in g,q,\bar q} \nonumber
p_j(T,\mu_j-U_j^{(mp)}) + \\  &&  +\sum_{j \in h} p_j(T,\mu_j-U_j^{(mp)})
-C(\rho^{(mp)})~,
\label{eqt8m} \\
&& \varepsilon^{MP}(T,\mu_j-U_j^{(mp)}) = \nonumber
\sum_{j \in g,q,\bar q} \varepsilon_j(T,\mu_j-U_j^{(mp)})+\\  && +\sum_{j
\in h} \varepsilon_j(T,\mu_j-U_j^{(mp)})+C(\rho^{(mp)})~,
\label{eqt10m}
\end{eqnarray}
where
\begin{eqnarray}
 C(\rho^{(mp)})&=& \frac{x \alpha}{\alpha +1 } \ (\rho^{(h)})^{\alpha +1 } -\nonumber \\&-&\rho^{(h)} \ F(\rho^{(h)})
  + \int_0^{\rho^{(h)}} F(t)dt - \nonumber \\ &-&
\frac{\gamma  {\cal A} }{1-\gamma} \left[ (\rho^{(mp)})^{1-\gamma}
- (\rho^{(h)})^{1-\gamma}\right]+\nonumber \\&+&\frac{\cal B}{4} \ \left[
(\rho^{(mp)})^{4/3}-(\eta \rho^{(h)})^{4/3}\right]~. \label{eqp}
\end{eqnarray}

\begin{figure}[thb]
\centerline{
\includegraphics[width=60mm,clip]{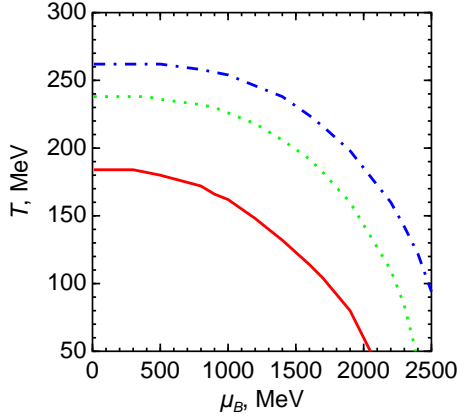}}
\caption{ The phase boundary calculated  in the MP  model (solid
line). The dotted and dot-dashed lines correspond to the state
where with the fraction of unbound quarks is 0.5 and 0.6,
respectively. }
 \label{boundmp}
\end{figure}

The MP model described above exhibits a crossover deconfinement
phase transition. The transition  temperature $T_c$ corresponds to
a maximum in the $T$-dependence of the heat capacity at the  given
value of $\mu_B$ (see the next Section). The resulting phase
boundary is shown in Fig. \ref{boundmp}.   At $T\lsim $ 50 MeV the
maximum of the heat capacity is not well defined. The calculation
in Fig. \ref{boundmp} was performed   with the physical values of
the parameters as introduced  in the next Section.

 In the MP  model hadrons survive at $T>T_c$. If the fraction
of unbound quarks is defined as
$\rho^{(pl)}/(\rho^{(pl)}+\rho^{(h)})$, then  one can see from
Fig. \ref{boundmp} that at $\mu_B=0$ and the temperature as high
as $T-T_c\sim 100$ MeV there is still   40$\%$  quarks that are
bounded inside  hadrons.

\section{The comparison with the lattice data}

The Lattice Gauge theory is the only approach that allows to
extract the physical EoS of QCD medium. To further constraint  the
phenomenological models for the EoS introduced in the last
Section, we will compare their predictions with the available LGT
results. We focus mainly on the  recent LGT findings  obtained in
(2+1)--flavor QCD at the finite temperature and chemical potential
~\cite{Fodor02,Fodor04}.

 In order to use the mixed phase and 2P models  for the further comparison
with lattice results one needs, however,  to take into account
that lattice calculations are generally performed with quark
masses heavier than those realized in the nature. Consequently,
the hadron mass spectrum generated on the lattice is modified.

In Refs.~\cite{Fodor02,Fodor04} the ratio of the pion mass $m_\pi$
to the mass of the $\rho$ meson is around 0.5-0.75, which is
roughly 3 times larger than its physical value. Thus, to compare
the model predictions with LGT results   the hadron mass spectrum
used in the model calculations should be properly scaled. For this
we use a phenomenological parametrization of the quark mass
dependence of the hadron masses $m_j(x)$ that was shown in
Refs.~\cite{KRT-1,KRT-2} to be consistent with the  MIT bag model
results as well as with LGT findings. For the non-strange hadrons
this parametrization reads \cite{KRT-1,KRT-2}:
\begin{eqnarray}
\label{non-str} \frac{m_j(x)}{\sqrt{\sigma}}\simeq \nu_{lj} a_1 x
+\frac{m_j / \sqrt{\sigma}}{1+a_2x+a_3x^2+a_4x^3+a_5x^4}~.
\end{eqnarray}
Here $x\equiv m_\pi / \sqrt{\sigma}, \ \nu_{lj}$ is the number of
light quarks inside the non-strange hadron ( i.e. $\nu_{lj}=2$ for
mesons and $\nu_{lj}=3$ for baryons) and $\sigma=(0.42 \ GeV)^2$.
 \begin{table}
  \caption{\label{tab:table2}  Parameters of the
    interpolation formulae (\ref{non-str})}
\centerline{
  \begin{tabular} {c c c c c c }
\hline
         $a_1$ & $a_2$& $a_3$& $a_4$&$a_5$ \\
%    \cline{2-3}
    \hline
      &   &  &  &  &    \\
   % 0.51$\pm$ 0.1 \ & $\frac{a_1\nu_{lj}\sqrt{\sigma}}{m_j}$ \ &0.115$\pm$ 0.02 \ &
  % -0.0223$\pm$ 0.008 \ & 0.0028$\pm$ 0.0015   \\
 0.51  & $\frac{a_1\nu_{lj}\sqrt{\sigma}}{m_j}$  &0.115 &
   -0.0223 & 0.0028  \\
      &   &  &  &  &    \\
\hline
 \end{tabular}}
\end{table}

 For strange hadrons that carries strangeness  $s_j=1$ and $s_j=2$
we have, respectively
\begin{eqnarray}
\label{str1}
 \frac{m_j(x)}{ \sqrt{\sigma}}&=&0.55 \nu_{lj} x+\frac{1.7\cdot 0.42 \
 \frac{m_j} {\sqrt{\sigma}}}{(1+0.068 x)}, \\ % \mbox{ for } s_j=1,\\
\frac{m_j(x)}{\sqrt{\sigma}}&=&0.5788 x +\frac{0.42  \frac{m_j}{\sqrt{\sigma}}}{(0.4758+0.0142
 x)}. % \mbox{for } s_j=2.
\label{str2}
\end{eqnarray}
Simultaneously with the change of the hadron mass spectrum with
the pion mass one needs to account for the shift of the transition
temperature $T_c$ with $m_\pi$. We use  the parametrization that
is extracted from LGT calculations ~\cite{KRT-1},
\begin{eqnarray}
\label{Tc}
\left(\frac{T_c}{\sqrt{\sigma}}\right)_{m_{\pi}/\sqrt{\sigma}}\simeq
0.4 +0.04(1) \ \left(\frac{m_{\pi}}{\sqrt{\sigma}}\right)~.
\end{eqnarray}

To compare our phenomenological model EoS with that obtained on
the  lattice in  Refs.~\cite{Fodor02,Fodor04} we use the modified
hadron mass spectrum form Eqs. (\ref{non-str})--(\ref{str2})
corresponding to   the pion  $m_\pi\simeq $ 508 MeV as fixed in
these LGT calculations. In the deconfined phase the current quark
masses and  gluon mass are in general also   free parameters. In
the present   calculations we fixed $m_u=m_d=65$ MeV, $m_s=2.08 \
m_u$ and $m_g\simeq $ 700 MeV as followed from the successful
description of the quark sector of the above LGT data in terms of
the quasi-particle model~\cite{Szabo03}.

  We look for a phase transition
at the appropriate temperature $T_c$ defined by Eq. (\ref{Tc}) by
varying mainly the bag constant $B$ in the 2P model or strength
parameter $\cal A$ in the mixed phase model. The further fine
tuning is carried out by means of remaining parameters (one
parameter in the 2P model and three ones for the mixed phase
model) to get the best description of LGT findings on temperature
dependence of different thermodynamical quantities.

 In the 2P
model, where the two phases do not interact with each other, the
critical temperature  $T_c$ is governed mainly by the value of the
bag constant $B$ and  the parameter $\alpha$ that characterizes
the  hardness of the EoS. Choosing $B^{1/4}=\rm 223 \ MeV$ and
$\alpha=2.1$ one gets $T_c \approx  176$ MeV and
$\varepsilon/T^4~|_{T_c}=7.84$  to be consistent with the LGT
results. We have to stress, however, that
 for some  values of the parameters, e.g.  for too heavy masses,  the set of
Eqs. (\ref{eqt12a})-- (\ref{eqt12c}) may have no solution.

\begin{figure}[thb]
\centerline{
\includegraphics[width=60mm,clip]{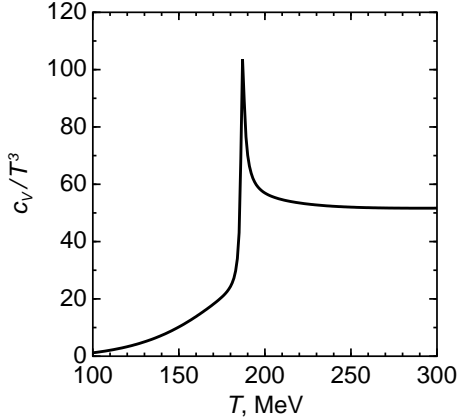}}
\caption{ Temperature dependence of the reduced heat capacity at
$\mu_B=0$ }
 \label{cV}
\end{figure}

In the MP model the critical temperature is defined at  the
position of the  maximum of the heat capacity
$$ c_V= \partial\varepsilon /\partial T|_{V=const} $$
The value of $T_c$  depends mainly on parameters that quantify the
quark/gluon interactions. With  ${\cal A}^{1/(3\gamma+1)}= 270$
MeV and $\gamma= 0.3$ the accepted value of the critical
temperature is seen in Fig. \ref{cV} to be 188 MeV. As was noted
in Ref. \cite{TNFNR03}, $\gamma=1/3$ corresponds to a string-like
quark interaction.

\begin{figure*}[thb]
\centerline{
\includegraphics[width=60mm,clip]{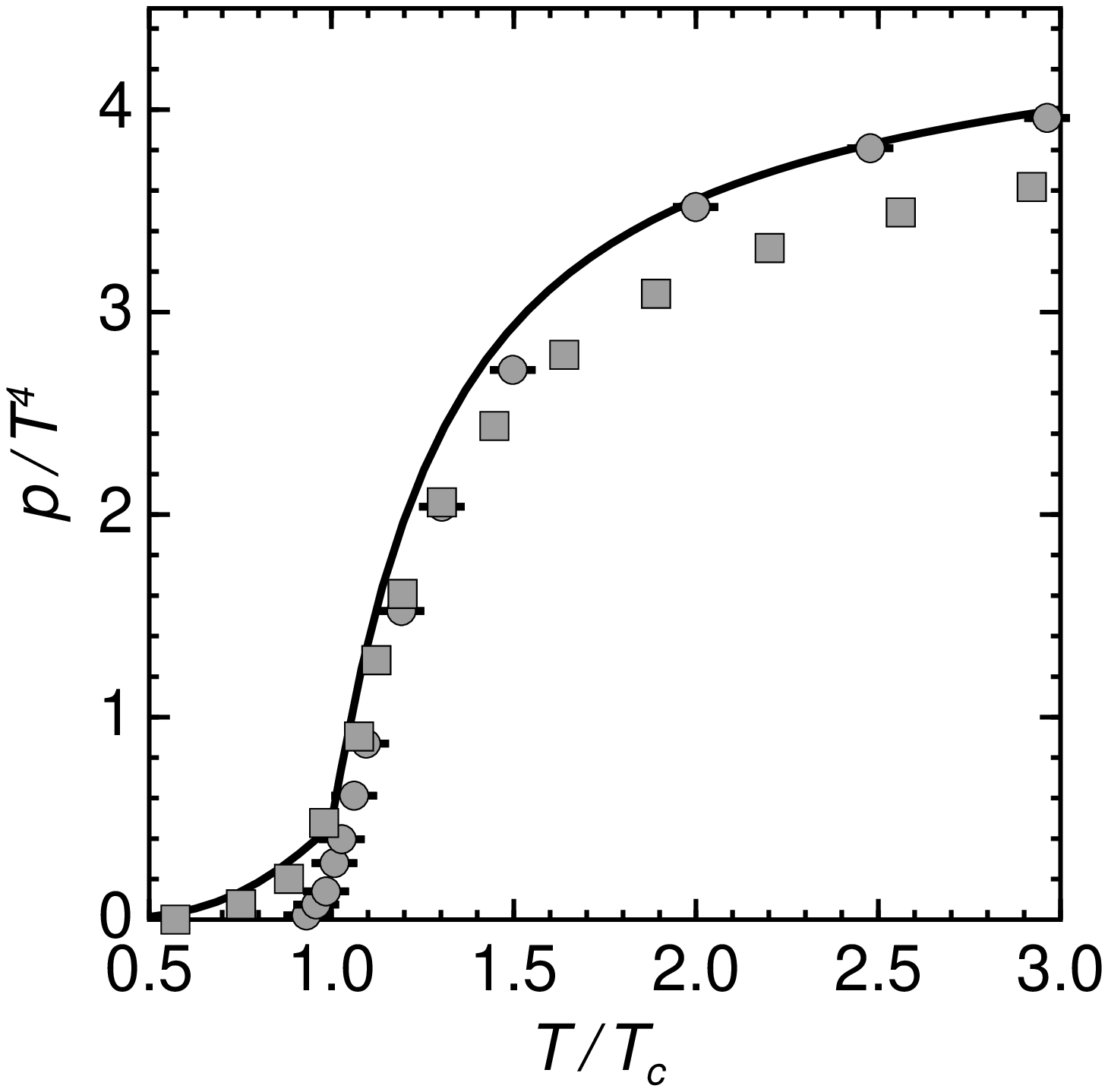} \hspace*{3mm}
\includegraphics[width=60mm,clip]{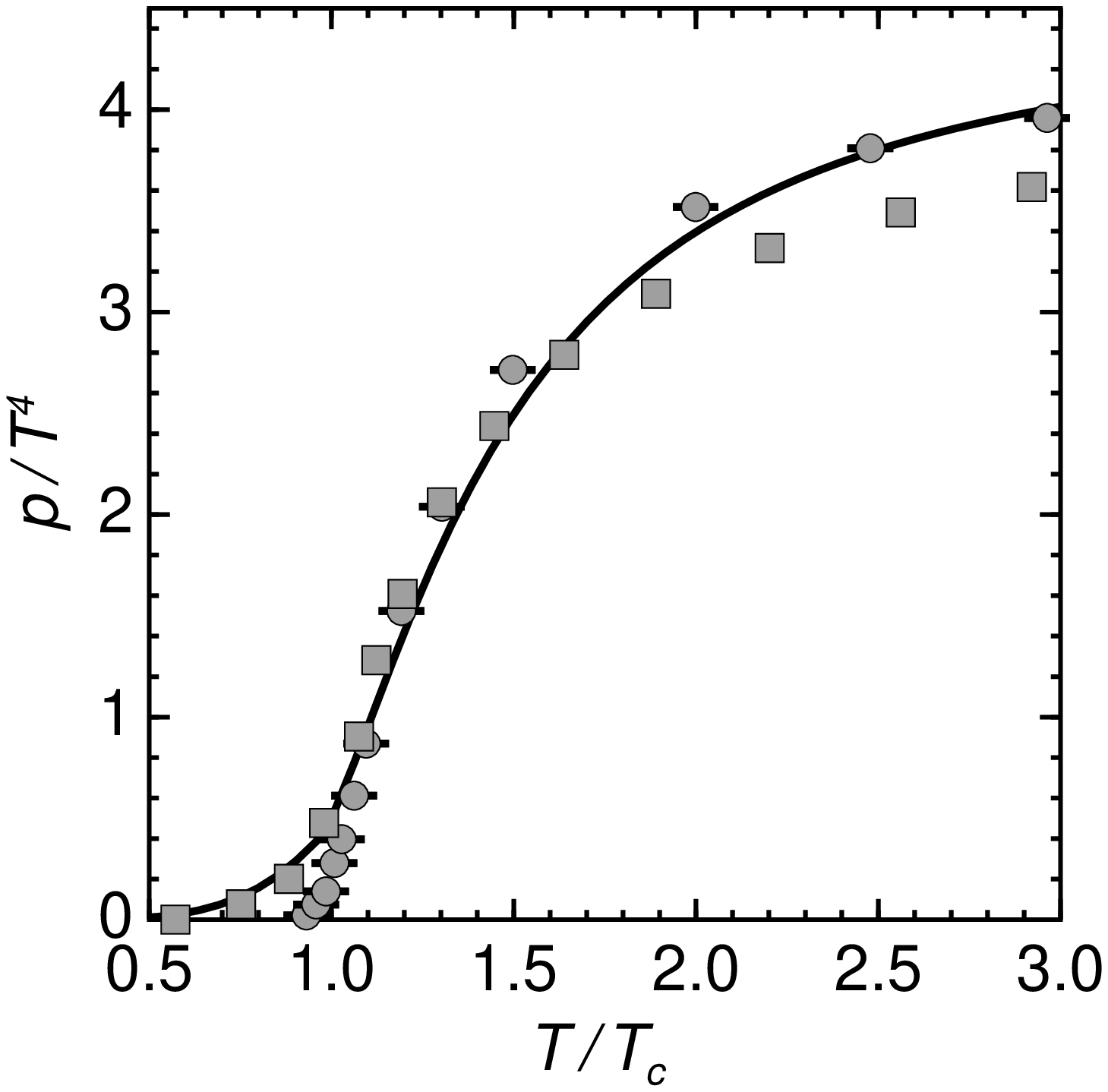}}
\caption{ The reduced pressure at $\mu_B=0$ in 2P (the left panel)
and MP (the right panel) models. Circles are the lattice data for
the (2+1)--flavor  QCD system~\cite{Fodor02,Fodor04} multiplied by
$c_0$, squires are the Bielefeld group data for the same case
\cite{KLP01} (see also results cited in~\cite{KRT-1}.)}
 \label{p0}
\end{figure*}

\begin{figure*}[thb]
\centerline{
\includegraphics[width=60mm,clip]{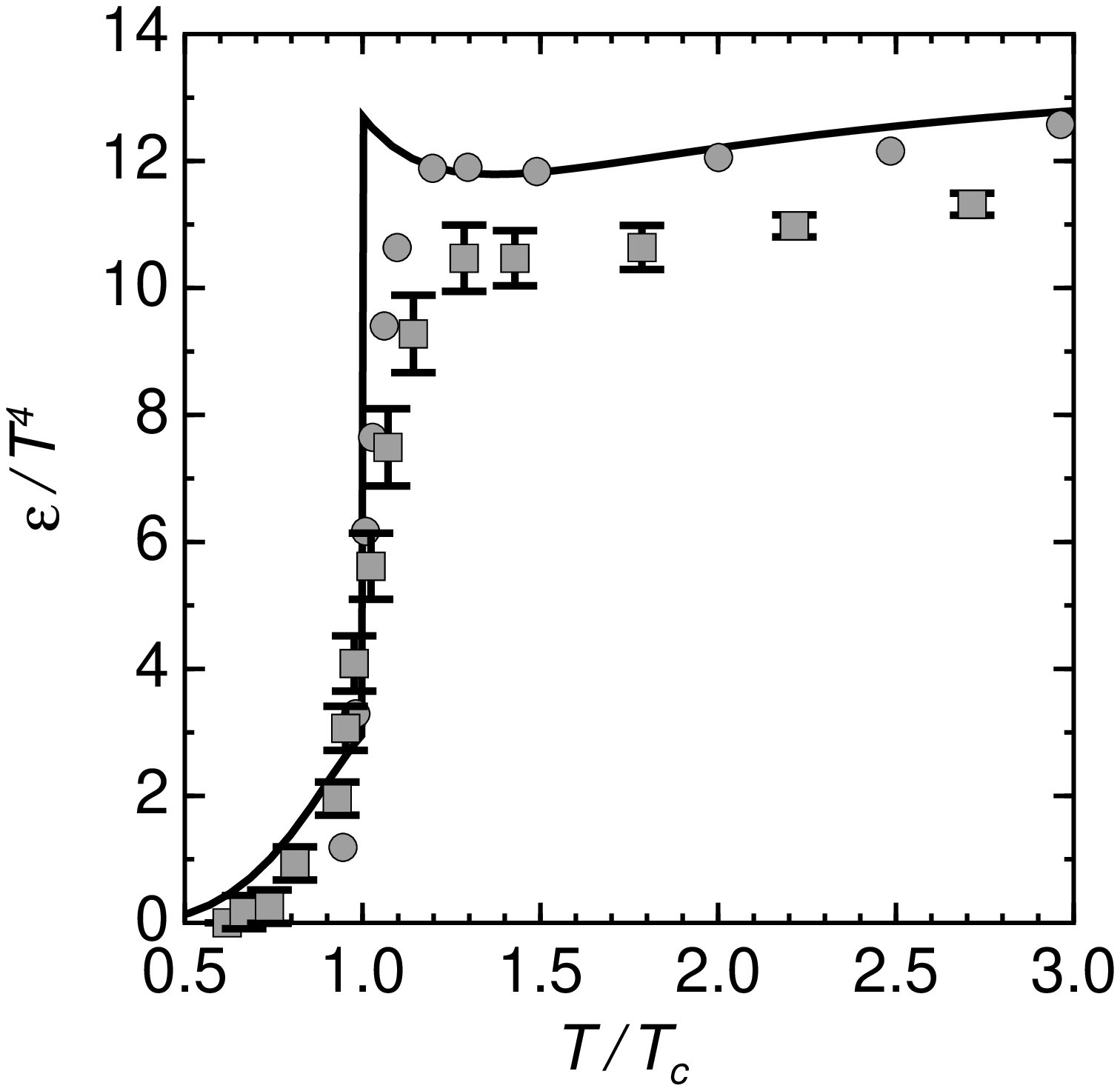} \hspace*{3mm}
\includegraphics[width=60mm,clip]{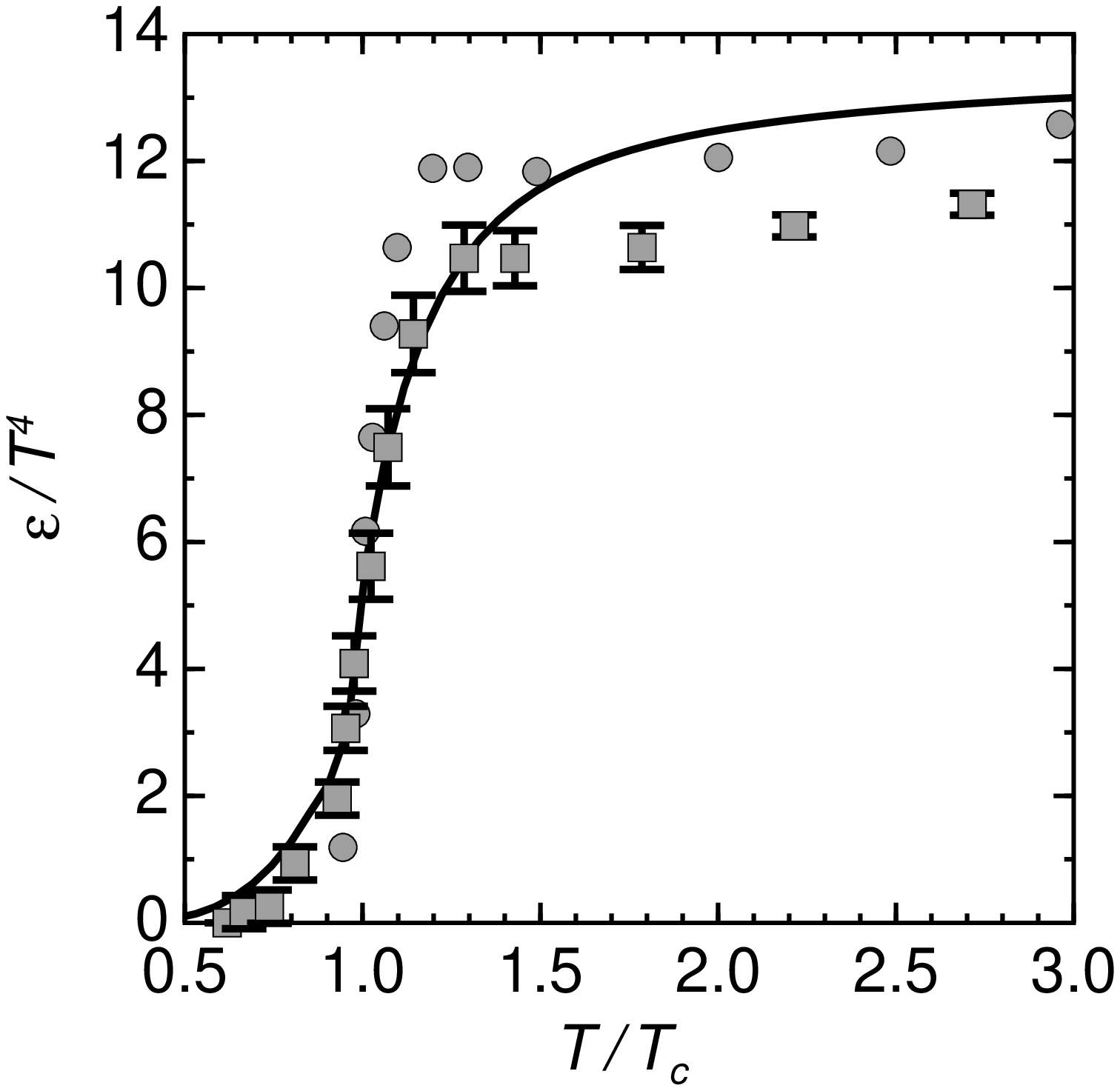}}
\caption{ The reduced energy density at $\mu_B=0$ in 2P (left
panel) and MP (right panel) models. Notation is the same as in
Fig. \ref{p0}}
 \label{e0}
\end{figure*}

In   Figs. \ref{p0} and \ref{e0} we show the comparison of the MP
and the 2P model predictions with LGT data obtained for the
thermodynamic pressure $p/T^4$ and the energy density $\varepsilon
/T^4$ at the  finite $T$ but for $\mu_B=0$.

The lattice calculations in Refs.~\cite{Fodor02,Fodor04} were done
on the lattices with $N_t=4$ temporal extension. To account for
the finite size effects,  the LGT results have to be  extrapolated
to the continuum limit corresponding to  $N_t \to \infty$. In
general, such a procedure  requires  a detailed LGT calculations
on the lattices with different $N_\tau$.  In
Refs.~\cite{Fodor02,Fodor04} to account approximately  for the
finite size effect the $N_\tau=4$ data for the basic thermodynamic
quantities were corrected being multiplied by the constant
factors: $c_0=0.518$ and $c_\mu=0.446$ for $\mu_B=0$ and
$\mu_B\neq 0$ respectively. These factors were  determined from
the ratios of the Stefan-Boltzmann ideal-gas limit for the
thermodynamic pressure to  its corresponding values calculated on
the lattice with $N_t=4$.

As it  is seen in Fig. \ref{p0}, a  smooth $T$--dependence of
pressure in a deconfined phase may be quite well reproduced within
both the MP and 2P  model.  However,  in the hadronic phase, that
is for $T/T_c<1$, the models overestimate LGT results from
Refs.~\cite{Fodor02,Fodor04}. In  Fig. \ref{p0}  also shown are
LGT results for   $(2+1)$--flavor QCD at $\mu_B=0$ from the
Bielefeld group~\cite{KLP01}. Improved gauge and staggered fermion
actions were used there on the lattices with temporal extent of
$N_t=4$ and $N_t=6$.  These  data were also extrapolated to the
chiral limit \cite{KLP01}. As seen  in Fig. \ref{p0}, the
Bielefeld  data exhibit a smaller limiting pressure as compared
to~\cite{Fodor02,Fodor04} and essentially higher pressure in the
hadronic sector, though the pion mass is $m_\pi=770$ MeV in the
latter calculations. Our models are seen in Fig. \ref{p0} to
coincide with Bielefeld results in the confined phase. The
strongly suppressed pressure at $T\leq T_c$ found in
Refs.~\cite{Fodor02,Fodor04} is non-physical and could be partly
related with too simplified procedure to extrapolate LGT results
to a continuum limit when applying the same constant scaling
factor for all values of temperatures.

 The energy density shown in  Fig. \ref{e0} behaves differently in the
 MP and 2P
model. As it is expected in the 2P model, that exhibits   the
first order phase transition, the $\varepsilon /T^4$ suffers a
jump at the critical temperature. This jump  corresponds to the
energy density change  by  $\Delta \varepsilon \sim 0.9 \ {\rm
GeV/fm^3}$. The LGT results on  the temperature dependence of
$\varepsilon/T^4$ are seen  in    Fig. \ref{e0}    to be
noticeably better reproduced within the MP than with the 2P model.
This is because the MP model exhibits a crossover type transition
as also found in the above LGT calculations. The difference
between LGT results obtained with an improved and standard action
is also seen on the level of  energy density.

\begin{figure*}[thb]
\centerline{
\includegraphics[width=60mm,clip]{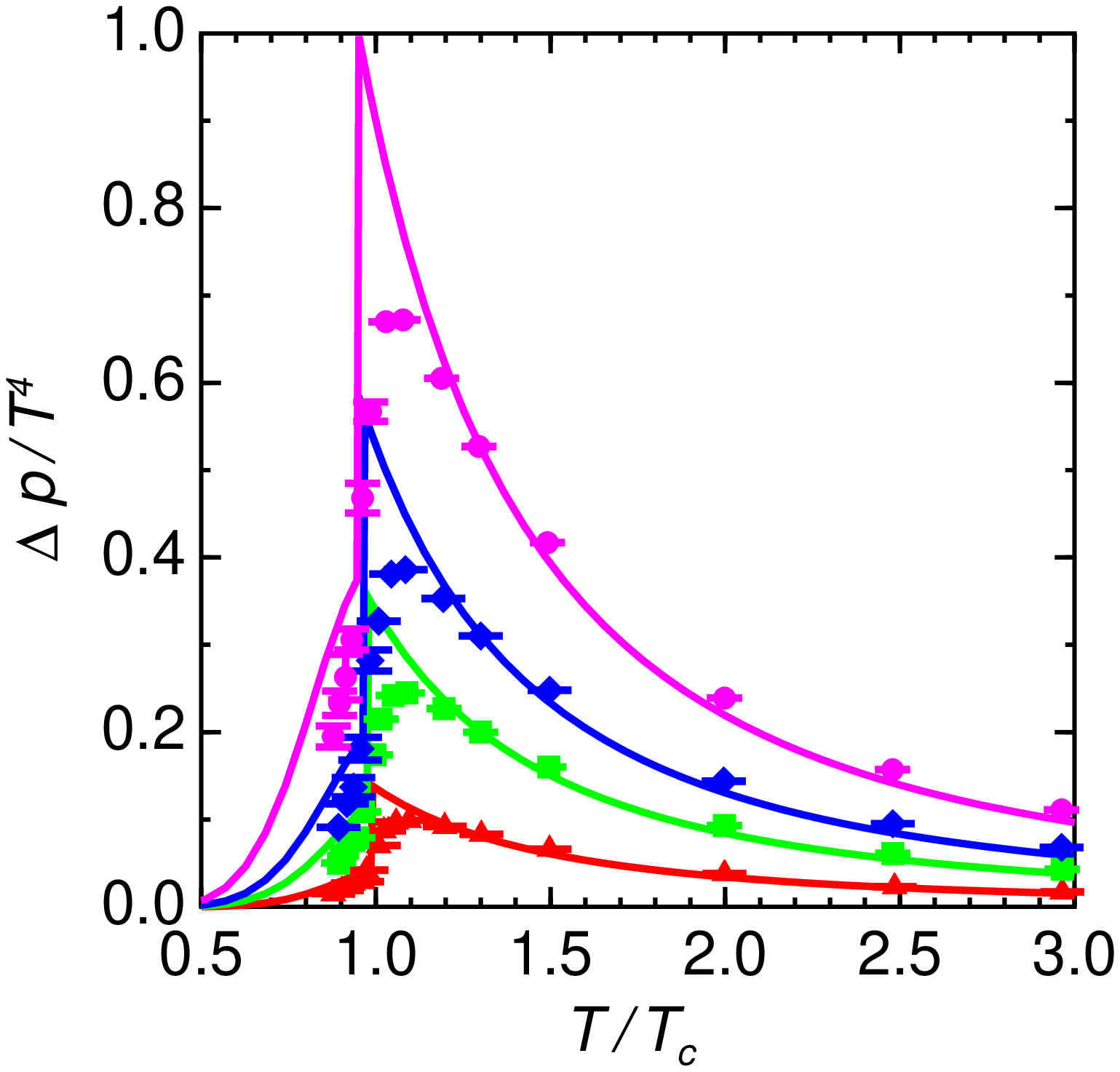} \hspace*{3mm}
\includegraphics[width=60mm,clip]{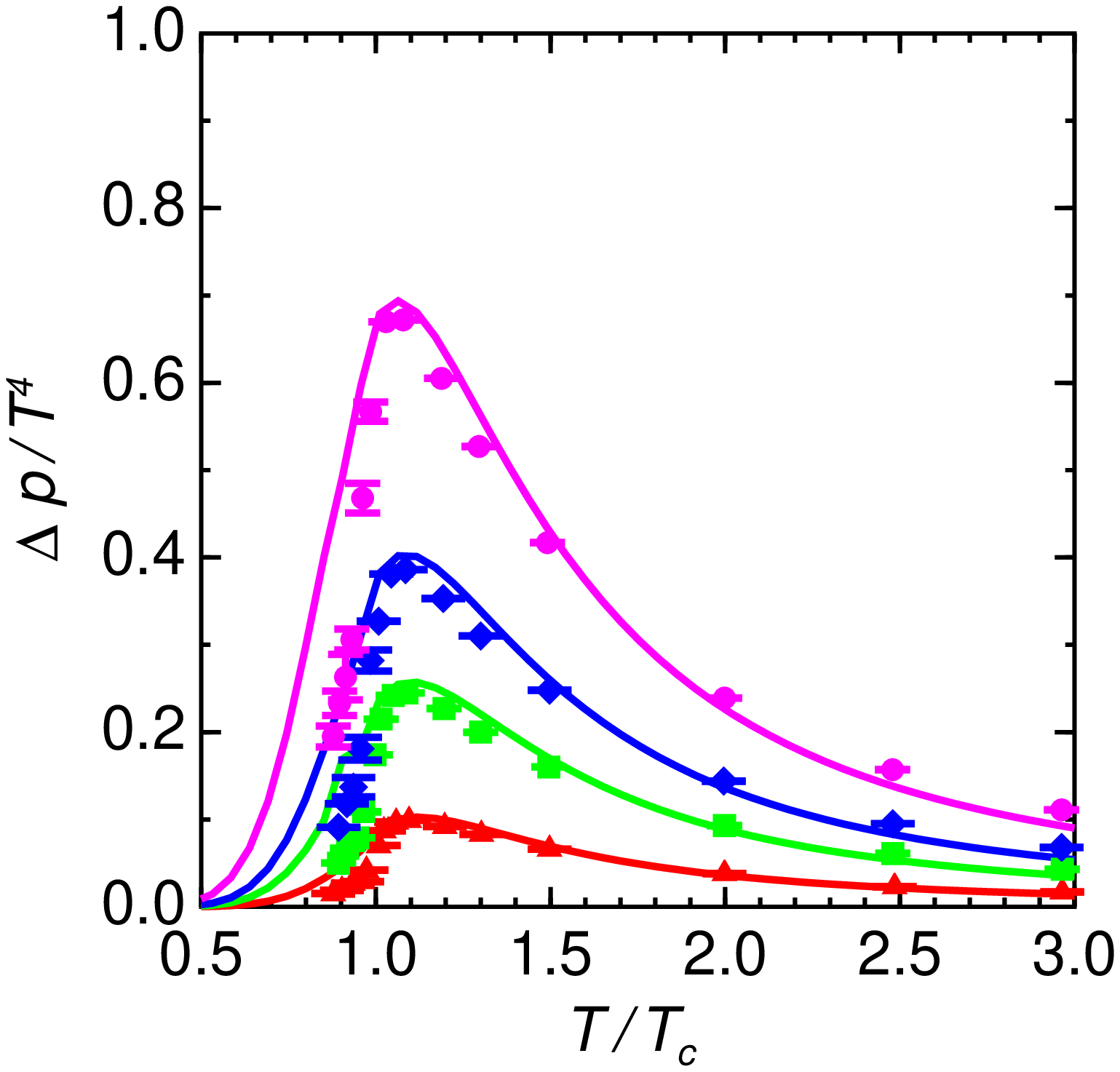}}
\caption{ Temperature dependence of the reduced pressure
$(p(\mu_B)-p(0))/T^4$ at the  baryon chemical potential $\mu_B=$
210, 330, 410 and 530 MeV  (from the bottom) within 2P ( the left
panel) and MP (the right panel) models. Points are lattice data
for the (2+1)--flavor system~\cite{Fodor02,Fodor04}  multiplied by
$c_\mu$. }
 \label{dpmu}
\end{figure*}

\begin{figure*}[thb]
\centerline{
\includegraphics[width=60mm,clip]{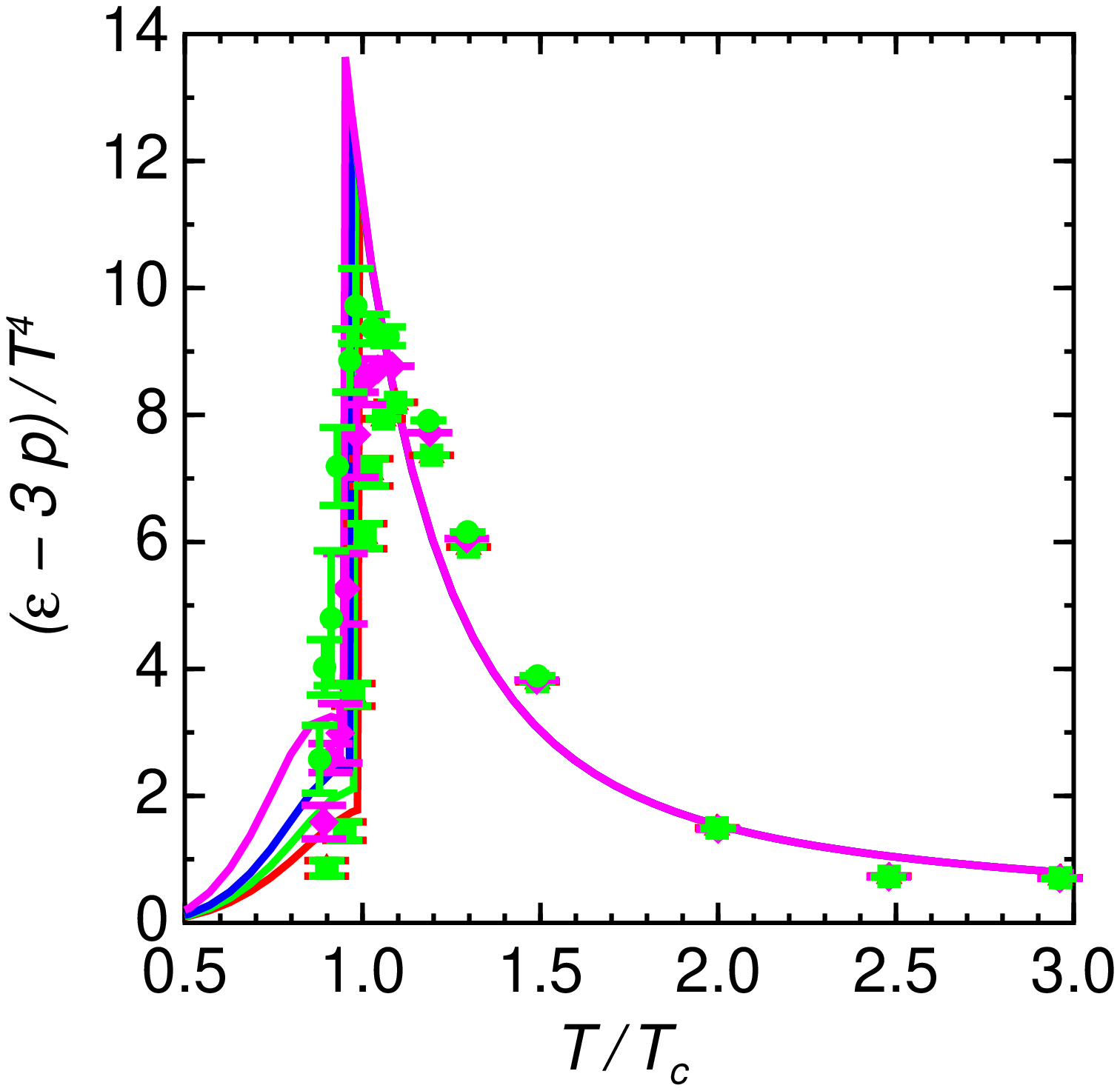} \hspace*{3mm}
\includegraphics[width=60mm,clip]{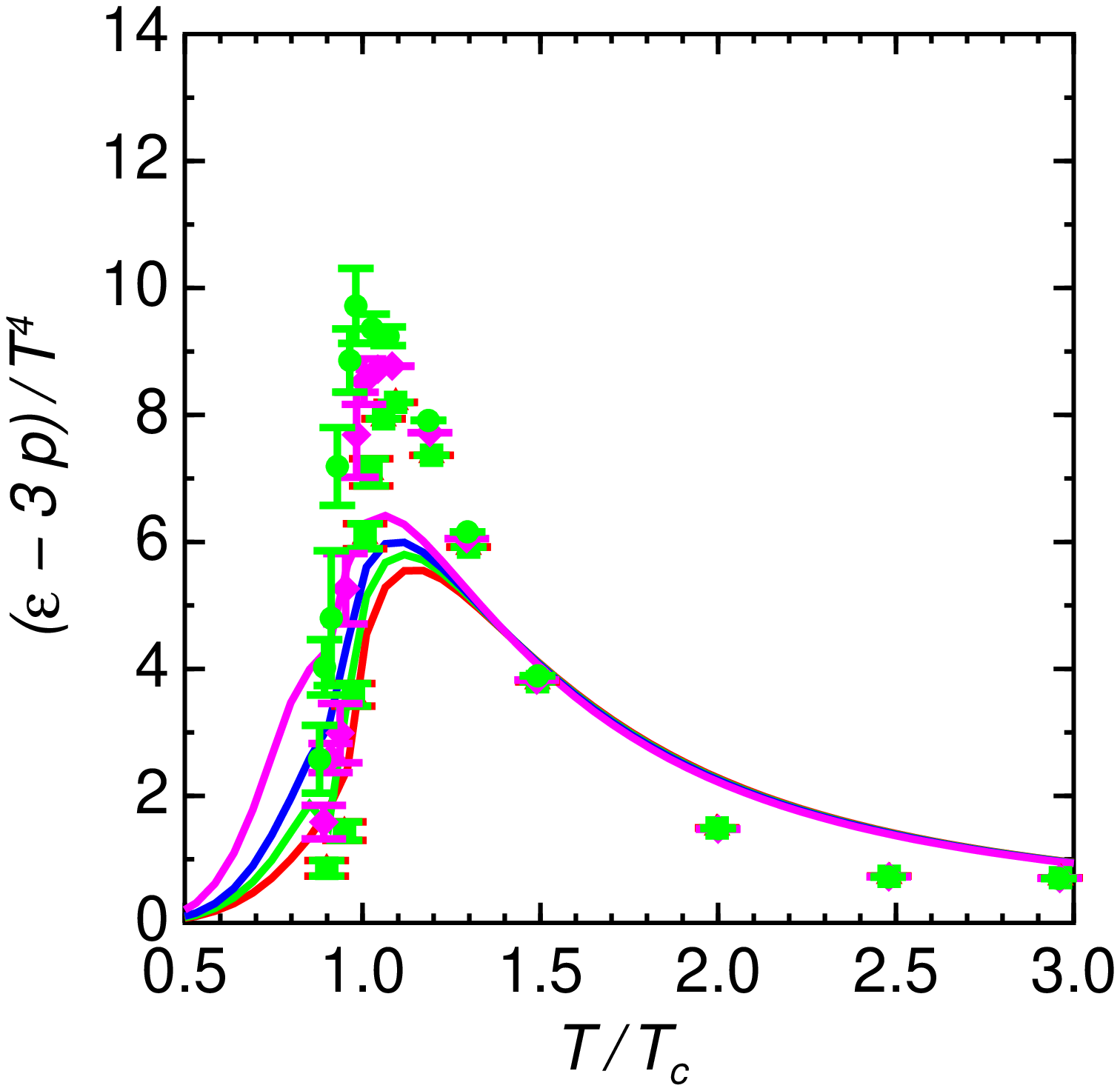}}
\caption{ Temperature dependence of the interaction measure
$(\varepsilon-3p)/T^4$ at the  baryon chemical potential $\mu_B=$
210, 330, 410 and 530 MeV (from the bottom) within 2P (the left
panel) and MP (the right panel) models. Points are lattice data
for the (2+1)--flavor  QCD system~\cite{Fodor02,Fodor04}
multiplied by $c_\mu$. }
 \label{demu}
\end{figure*}

Having established the model parameters at $\mu_B=0$ we can
further study the model comparisons with LGT results at the finite
baryon density. The temperature dependence of pressure and energy
density for finite values of $\mu_B$ is shown in Figs. \ref{dpmu}
and \ref{demu} in terms of the "net bryonic pressure" $\Delta p
/T^4= (p(T,\mu_B)-p(T,\mu_B=0))/T^4$ and the "interaction measure"
$\Delta  /T^4= (\varepsilon -3p)/T^4$.

The $T$ dependence of  $\Delta p /T^4$ for different values of
$\mu_B$ is quite well reproduced by both  the MP and 2P model. The
fall  of $\Delta p /T^4$ for $T\geq T_c$  is entirely determined
by the value of the coupling  $g$ that describes the strength of
the interactions of quasi-particles and their effective mass. The
observed fall  does not require  any artificial reduction of the
number of quark--gluon degrees of freedom. It turns out that in
both models a similar value of $g=0.5$ is necessary to reproduce
LGT results.

The interaction measure $\Delta /T^4$ exhibits a rather sharp
maximum \ \ slightly\ \  above $T_c$ with\ \  the shape of $T$-dependence that
is  weakly changing  with $\mu_B$. In general, both models
reproduce the above  properties  of the interaction measure.
However, quantitatively the  $\Delta /T^4$ is overestimated in the
2P model and underestimated in the MP model near  the maximum.

The interaction measure characterizes  the strength of
interactions in a system. It is  equal to zero for the EoS of the
ultrarelativistic ideal gas of massless particles where
$\varepsilon=3p$. In the considered  2P model and at $T>T_c$ we
are dealing with a  gas of massive quarks and gluons interacting
via the HTL-like potential (\ref{U2P}). In contrast, in the MP
model at $T>T_c$,  there are interacting unbound quarks, gluons
and bound quarks  within hadrons. The fraction of the bound quarks
amounts in about 85 $\%$ at $T\sim T_c$ (which  allows  to
describe this region in terms of the resonance gas
model~\cite{KRT-1,KRT-2}) and almost vanishes at $3T_c$ ($\sim
5\%$). In this  context the quark--gluon plasma   may be
considered as a strongly interacting correlated system
~\cite{SZ03}. In the confined phase there is   an admixture of
quarks at $T<T_c$ until about $0.9~T_c$. This property of the
model  is very essential for a possible explanation of the "horn"
structure in the $K^+/\pi^+$ excitation function~\cite{mg:04} due
to manifestation of the strangeness distillation effect near the
critical end point~\cite{TP05}.

\begin{figure*}[thb]
\centerline{
\includegraphics[width=60mm,clip]{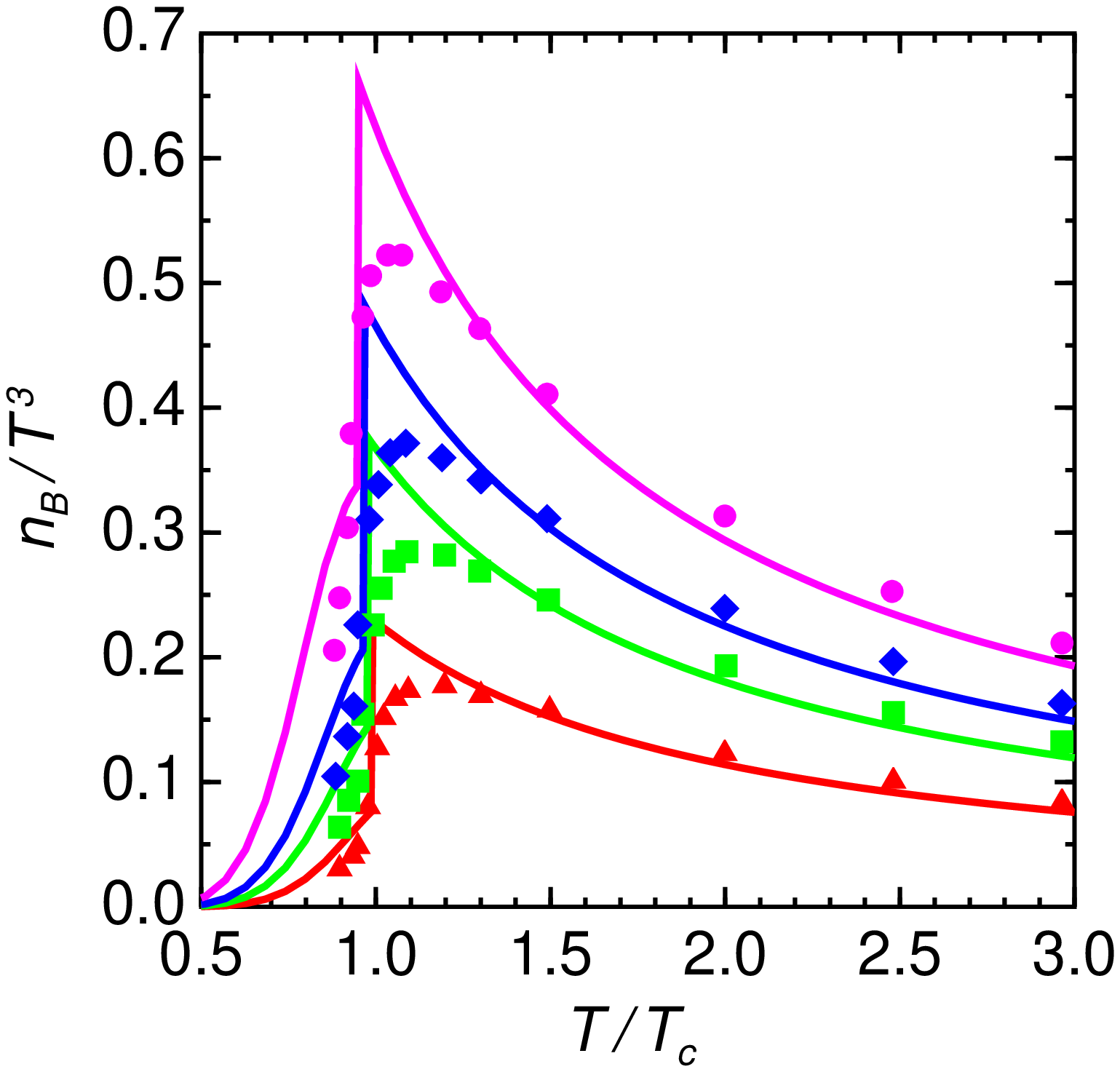} \hspace*{3mm}
\includegraphics[width=60mm,clip]{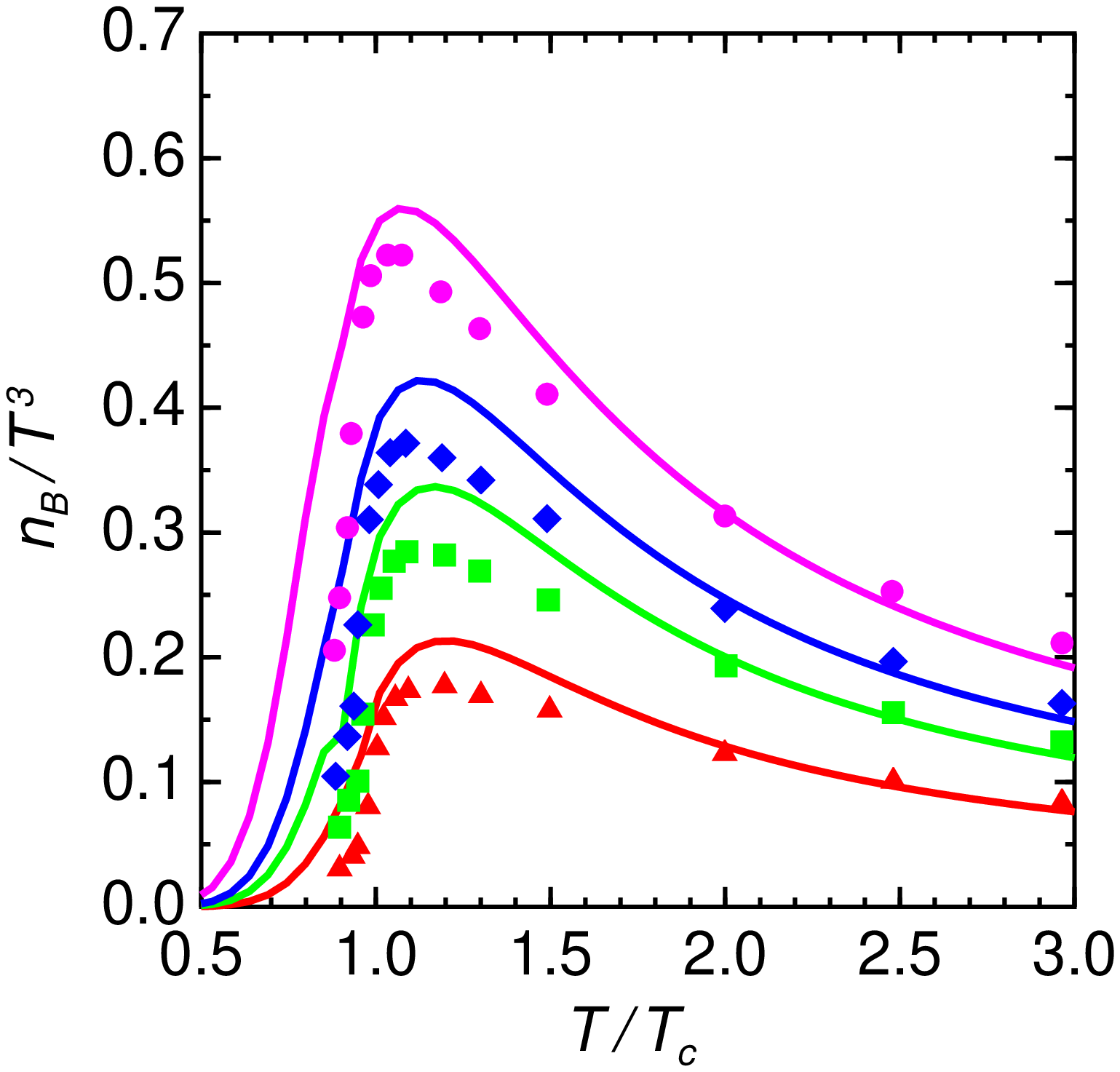}}
\caption{ Temperature dependence of the baryon density at the
baryon chemical potential $\mu_B=$~ 210, 330, 410 and 530 MeV
(from the bottom) in 2P (the left panel) and MP (the right panel)
models. Points are lattice data for the (2+1)--flavor QCD
system~\cite{Fodor02,Fodor04} multiplied by $c_\mu$. }
 \label{nb}
\end{figure*}

The\ \ model comparison with \ \  LGT results for the baryon density is
shown in Fig. \ref{nb}. It is clear from this figure that in the
hadronic sector the baryon density  $n_B/T^3$ obtained on the
lattice is smaller than that predicted by the models. However,
above $T_c$ there is quite a good agreement of the model with LGT
results. This is particularly a case for the MP model which shows
a better description of LGT data near the phase transition.

In our models the absolute values of  $\Delta p /T^4$, $\Delta
/T^4$ and $n_B /T^3$  are strongly affected by the parameter
$\eta$ appearing in Eq. (\ref{rhomp}). In the actual calculations
the $\eta=0.025$, thus  it is essentially smaller than $\eta=1$
that was found  in our earlier parametrization based only on the
LGT  data obtained for  $\mu_B=0$ ~\cite{TNFNR03}. If $\eta=1$ is
to  be  substituted in our actual calculations, then all the above
quantities would increase by a factor of two.

The properties and the behavior of the LGT thermodynamics at
finite T and $\mu_B$   have been recently discussed in the context
of the Polyakov-loop-extended Nambu and Jona-Lasinio (PNJL)
model~\cite{Weise05}. This PNJL model represents a minimal
synthesis of the spontaneous chiral symmetry breaking and
confinement. The model correctly describes the pion properties but
obviously is not applicable near  the nuclear ground state. It
also does not contain the resonance contributions to the QCD
thermodynamics nor the hadronic correlations below and above $T_c$
that are essential near the phase transition. Nevertheless,  the
PNJL model reproduces the LGT data  \cite{Allton} obtained in 2
flavor QCD on the pressure difference and the quark number density
at various temperatures and  chemical potentials remarkably well
\cite{Weise05}. However, in the PNJL model the interaction measure
$\Delta / T^4$ was found to be underestimated by $\sim 25\%$
similarly  as seen in Fig.\ref{demu} from our MP model comparison
with the (2+1)--flavor QCD results obtained in LGT.

\begin{figure}[thb]
\centerline{
\includegraphics[width=70mm,clip]{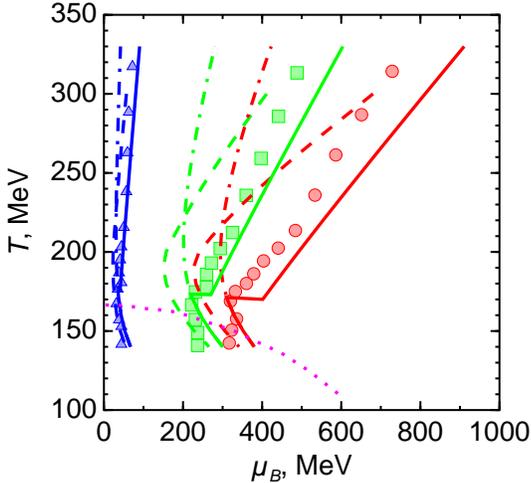} \hspace*{3mm}}
\caption{ Phase trajectories in the $T-\mu_B$ representation.
Circles, squares and triangles are the lattice 2-flavor QCD
results~\cite{EKLS05}  for $S/N_B=$ 30, 45 and 300, respectively.
 The (2+1)--flavor model predictions are plotted by solid (2P EoS),
 dashed (MP EoS) and dot-dashed
 (Hadronic EoS) lines for every value of the reduced entropy. The dotted
 line parameterizes the freeze out curve~\cite{CR98}.
}
 \label{lat_tr}
\end{figure}

The phenomenological models considered here describe  the EoS of
the QCD matter in a  broad  range of thermal parameters that
includes the hadronic and quark gluon plasma phase. These models
are also applicable in cold nuclear matter as they satisfy
essential phenomenological constraints expected near nuclear
saturation.  In heavy ion collisions, dense QCD matter created in
the initial stage  is expected to thermalize and expand without
further generation of the entropy $S$. In the realistic expansion
scenario some particles may be crated and/or absorbed implying
changes in the total entropy of the  system. In general, it is
more convenient to consider the EoS at  fixed entropy per baryon
($S/N_B$). This thermodynamic quantity should be  strictly
conserved in an equilibrium case and is also less affected by any
possible particle losses or creation during the expansion stage.
The predictions of our  models for the evolution path in the
$(T,\mu_B)$-plane as obtained from the condition of fixed $S/N_B$
is shown in  Fig. \ref{lat_tr}. There are  still no such
isentropic lattice data for  (2+1)--flavor system. Recently, the
isentropic EoS was obtained on the lattice for 2-flavor QCD at
finite $\mu_B$ \cite{EKLS05}, however still for  non--physical
mass spectrum that corresponds to the pion mass $m_\pi\simeq 770$
MeV. These data are plotted in Fig. \ref{lat_tr} together with our
model results obtained with the EoS parameters that are fixed for
$m_\pi\simeq 508$ MeV and for (2+1)--flavor system.

In general, in the high-temperature deconfined phase,   one should
not expect a large difference between  2 and (2+1)--flavor
thermodynamics. The value \ \  of the \ \ quark mass in the quark--gluon
plasma is also not relevant thermodynamically if $m_q/T<1$. In the
 hadronic phase the number of quark flavors is as well not
essential and leads only to a moderate change of the global
thermodynamics. However, here the value  of the quark/pion  mass
is of particular importance as it influences the hadronic mass
spectrum. Due to the non-physical and still large  pion mass used
in the actual LGT calculations  it is not straightforward to
associate the values of the reduced entropy with the specific
bombarding energy. In particular, as noted in Ref.~\cite{EKLS05},
the correspondence of $S/N_B=$~ 30, 45 and 300 to the AGS, SPS and
RHIC energies, respectively, is only a rough  approximation.  The
QGSM transport model results~\cite{ST06} for central Pb+Pb
collisions at the top SPS energy show that the isentropic regime
is reached after about 1 fm/c with  $S/N_B\approx$ 25. Also
calculations performed in terms of  3--fluid relativistic
hydrodynamic model show that the isentropic expansion of central
Pb+Pb collisions at the bombarding energy 158 and 30 AGeV results
in $S/N_B\approx$ 30 and 15~\cite{IRT05} respectively. Thus, the
above  dynamical models imply  noticeably lower values of $S/N_B$
than that obtained within actual LGT calculations \cite{EKLS05}.
The main origin of the above differences is related with still too
large quark mass used on the lattice.

As  seen in  Fig. \ref{lat_tr} the MP and 2P models  reproduce the
general trend of the lattice trajectories.  The lattice evolution
paths  are just between  2P and MP model predictions. With
increasing   $S/N_B$ these differences are noticeable smaller. The
hadronic EoS predicts higher initial temperatures, however all
three phenomenological models give similar results for the
freeze-out temperature. It is interesting to see in Fig.
\ref{lat_tr}  that irregularity appearing  near the turning point
of the lattice trajectory correlates with the flattening of the
T-dependence in the Gibbs mixed phase resulting in the  2P model.

The phenomenological model results,  discussed so far, were
obtained assuming the  hadronic mass spectrum that corresponds to
the pion mass $m_\pi=508$ MeV. The extrapolation of  the  EoS to
the physical limit is  quite straightforward. It amounts in
replacing the $m_j(m_\pi)$ masses by their physical values. The
quark and gluon masses  are kept to be the same as being extracted
from  the LGT data. This approximation is justified  since the
change of $m_q$ in the interval $5<m_q<70$ MeV does not influence
  the thermodynamics  in the plasma phase~\cite{Szabo03} very much. Clearly,
taking the physical limit in the EoS  also requires to account for
the shift  in   $T_c$. In the 2P model the critical temperature is
recalculated according to Eq. (\ref{Tc}) and fitted in the model
by the bag $B$ constant  and the coupling $g$ to  satisfy also the
condition that the critical energy density $\varepsilon_c /T_c^4
\simeq 6 \pm 2$~ as found in LGT \cite{KRT-1}. Within the 2P model
the physical limit is achieved choosing: $B^{1/4}=\rm 207 \ MeV$
and $g=0.7$ which results in $T_c= 173.3$ MeV and
$\varepsilon_c/T_c^4=7.83$.

The extrapolation of  the MP model EoS to the physical limit is
less transparent due to a rather strong
 nonlinear relation between the hadronic  and plasma phase. In
this model the physical  limit is approximately accounted for by
replacing  the LGT mass spectrum by  its   physical form. All
further parameters that are require in the MP model to quantify
the EoS are kept  the same  as that found in the comparison of the
model predictions with the LGT results. With the above chosen
model parameters the  crossover deconfinement transition appears
at  $T_c=183$ MeV. Note that  the phase boundaries  in the
preceding Section were calculated for these physical parameters of
the EoS.

\section{Summary}

 We have
formulated  two different  phenomenological models for the
equation of state within  the quasi-particle approximation of the
QCD matter: the \ \  two\ \  phase (2P) model with the first order
deconfinement transition and the mixed phase (MP) model in which
transition  from hadronic phase to quark--gluon plasma is of the
crossover type.

In our approach both the hadronic and the quark--gluon plasma
phase are considered to be the non-ideal systems. The interactions
between constituents are included within the mean-field
approximation. The modified mean--field Zimanyi model is applied
to describe  the interacting resonance gas component. In this
approach, the saturation properties of a symmetric nuclear matter
in the ground state are reproduced correctly and the Danielewicz
constraints resulting from heavy--ion collision data  at
intermediate energies are well fulfilled.

The \ \  quark--gluon  phase in the 2P model is constructed as a
massive quasi-particle system supplemented by the
\ \ density-dependent potential term which simulates the HTL
interactions. The\ \  first\ \  order\ \   phase transition from the hadronic
phase to the deconfined quark--gluon plasma is constructed within
the 2P model  by means of the Gibbs phase equilibrium conditions.

In  the MP model the   coexistence and  correlations between
quarks/qluons and hadrons are  assumed near  deconfinement. In
addition to the HTL-like interaction term  a string-like
interaction is introduced between both unbound quarks/qluons and
quarks that are confined within hadrons. In this model  we are
dealing with strongly interacting QCD matter which exhibits  a
crossover-type deconfinement phase transition.

The models are constructed in the way  to be thermodynamically
consistent and to reproduce the properties of the EoS as
calculated on the lattice.  The limited set of model parameters is
defined from the constraints imposed by the recent lattice data on
 the temperature and chemical potential dependence of the basic
thermodynamical observables. The  comparison of the model
predictions with LGT data was performed within the same set of
approximations as used on the lattice. Of particular importance is
a correct treatment of the hadronic mass spectrum which in the LGT
calculations is  non-physical due to the still too large value of
the quark mass.

Keeping in mind a principal difference between the  first order
and crossover type phase transition both the 2P and MP model were
shown to provide quite satisfactory description of the LGT
thermodynamics for (2+1)--flavor QCD. Both models reproduce the
$T$ and $\mu_B$ dependence of the main thermodynamic quantities in
a broad  range of thermal parameters. The observed  deviations of
the model predictions  from  the lattice results  near $T_c$ and
in the hadronic sector for (2+1)--flavor case may  be, to a large
extend, attributed to uncertainties in the LGT data due to the
finite size effect. The predicted isentropic trajectories in the
phase diagram were  shown  to be  consistent  with that recently
calculated on the lattice  within  the 2-flavor QCD.

The phenomenological equations of state constructed here satisfy
all physically relevant constraints  expected in the  cold  and
excited nuclear matter. These EoS can be applied in a broad
parameter range that  covers the region of deconfinement
transition in QCD. Thus, both the MP and 2P EoS could  be used as
an input in dynamical models that describe the space-time dynamics
and evolution  of a medium created in heavy ion collisions. Within
hydrodynamic models our EoS  can be important to study  the role
and influence of   deconfinement and the order of the phase
transition on physical observables.  Such  studies  are in
progress.

\vspace*{5mm} {\bf Acknowledgements} \vspace*{5mm}

We are grateful to D. B.~Blaschke, Yu. B.~Ivanov, V. N.~Russkikh,
S.A.~Sorin and D. N.~Voskresensky for interesting  discussions and
comments. This work was supported in part by the Deutsche
Forschungsgemeinschaft (DFG project 436 RUS 113/558/0-3), the
Russian Foundation for Basic Research (RFBR grants 06-02-04001 and
05-02-17695) by the special program of the Ministry of Education
and Science of the Russian Federation (grant RNP.2.1.1.5409) and
by the Polish Committee for Scientific Research (KBN-2P03B 03018).

\end{document}